\documentclass[10pt]{iopart}

\usepackage{xcolor}
\usepackage{graphicx}
\usepackage{dcolumn}
\usepackage{bm}
\usepackage{booktabs}    
\usepackage{multirow}     
\usepackage{subfigure}
\usepackage[para,online,flushleft]{threeparttable}
\usepackage{textcomp}
\usepackage[utf8]{inputenc}
\usepackage[T1]{fontenc}
\usepackage{mathptmx}
\usepackage{rotating}
\usepackage{pdflscape,caption}
\usepackage{cite}

\usepackage{iopams}
\expandafter\let\csname equation*\endcsname\relax
\expandafter\let\csname endequation*\endcsname\relax
\usepackage{amsmath}

\newcommand{\ANL}{Materials Science Division, Argonne National Laboratory, Argonne, Illinois 60439, USA}
\newcommand{\LANL}{Theoretical Division, Los Alamos National Laboratory, Los Alamos, New Mexico 87545, USA}
\newcommand{\CEA}{Univ.\ Grenoble Alpes, CEA, INAC-SP2M, L\_Sim, F-38000, Grenoble, France}
\newcommand{\ICL}{Department of Materials, Imperial College London, London SW7 2AZ, United Kingdom}
\newcommand{\UofC}{James Franck Institute, University of Chicago, Chicago, Illinois 60637, United States}
\newcommand{\UIC}{Department of Physics, University of Illinois at Chicago, Chicago, Illinois 60607, USA}

\begin{document}

\ioptwocol

\twocolumn[
\begin{@twocolumnfalse}
\title{Exploring Metastable States in UO$_2$ using Hybrid Functionals and Dynamical Mean Field Theory}

\author{Laura E.\ Ratcliff}     
\ead{laura.ratcliff08@imperial.ac.uk}
\address{\ICL} 
\author{Luigi Genovese}
\address{\CEA}
\author{Hyowon Park} 
\address{\ANL}\address{\UIC}
\author{Peter B. Littlewood}     
\address{\ANL}\address{\UofC} 
\author{Alejandro Lopez-Bezanilla}
\address{\LANL}

\date{\today}

\begin{abstract}
A detailed exploration of the $f$-atomic orbital occupancy space for UO$_2$ is performed using a first principles approach based on density functional theory (DFT), employing a full hybrid functional within a systematic basis set.
Specifically, the PBE0 functional is combined with an occupancy biasing scheme implemented in a wavelet-based algorithm which is adapted to large supercells. The results are compared with previous DFT+U calculations reported in the literature, while dynamical mean field theory (DMFT) is also performed to provide a further base for comparison. This work shows that the computational complexity of the energy landscape of a correlated $f$-electron oxide is much richer than has previously been demonstrated.  The resulting calculations provide evidence of the existence of multiple previously unexplored metastable electronic states of UO$_2$, including those with energies which are lower than previously reported ground states.  
\end{abstract}
\end{@twocolumnfalse}
]

\section{Introduction}
Decades after efforts were devoted to unveiling the physical properties of UO$_2$, this f-electron oxide remains an active and exciting material of study in condensed matter physics.
Uranium is the most studied of the actinide elements, but its successful industrial use as a fissile material has not run in parallel to the refinement of computational tools suitable for computational modeling of its properties. 

\begin{figure}
\centering
\includegraphics[width=0.25\textwidth]{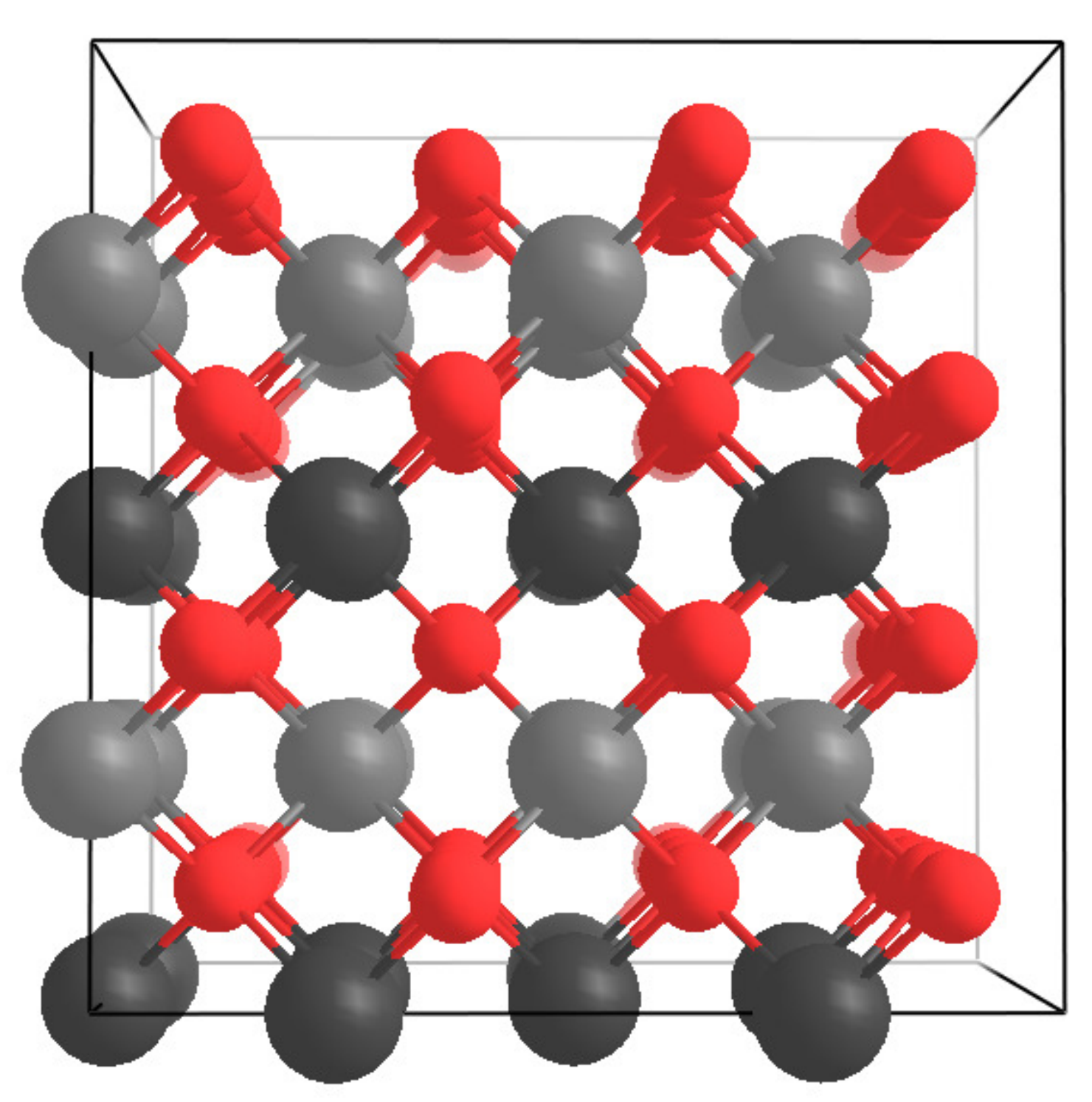}
\caption{Depiction of the 96 atom ($2 \times 2 \times 2$) fluorite-type supercell of UO$_2$, where U atoms with spin up (down) are depicted in dark (light) grey and O atoms are in red.}\label{fig:uo2_struc}
\end{figure}

The oxide obtained upon combining U with O in a tetravalent bonding, UO$_2$, is the most widespread uranium mineral. Further oxidation of UO$_2$ leads to a hexavalent atom of U in the form of UO$_2^{2+}$. In the crystalline form, fully stoichiometric UO$_2$ exhibits the fluorite-structure (Fm$\bar{3}$m), where the cations are distributed in a face centered cubic structure and the anions are located at tetrahedral sites, yielding a simple cubic sublattice within the unit cell (see Fig.~\ref{fig:uo2_struc}). UO$_2$ is a type-I traverse antiferromagnetic (AFM) Mott insulator exhibiting strong correlations between 5$f$-electrons and $f$-$f$ transitions across a band gap of $\sim$2~eV. The ground state of the tetravalent U ion in UO$_2$ (the oxidation state is $+4$) exhibits a population of 2 electrons in the seven-fold degenerate 5$f$-orbitals. 
As with the other first elements of the actinide series, the strongly correlated U 5$f$ states are at the boundary between simple localized and itinerant pictures of electrons, with $f$-orbitals partially occupied with electrons strongly localized close to the atomic core, and yielding a small crystal field splitting. The shielding of the U 5$f$-orbitals by the $s$, $p$, and $d$-orbitals in UO$_2$ is not as large as in the lanthanide series, which allows valence electrons to interact weakly with the O $p$-orbitals forming the ligand environment.  

Neutron scattering experiments of UO$_2$ under an external magnetic field suggest that below the N\'eel temperature the O sublattice undergoes a shear deformation. This spin-lattice coupling induces a non-collinear 3-\textbf{k} magnetic structure, which preserves the cubic symmetry with magnetic moments and oxygen atom displacements along the $<111>$ direction \cite{Faber1975,Burlet1986}.

A number of numerical methods have been used to model UO$_2$, ranging from density functional theory (DFT)~\cite{Dorado2009,Dorado2013,Krack2015} to dynamical mean field theory (DMFT)~\cite{Amadon2012,Yin2011} and self-interaction correction (SIC)~\cite{Petit2002}. However the most accurate method to describe both the crystal and defected configurations remains controversial. 
DFT~\cite{Hohenberg1964,Kohn1965} calculations with exchange correlation functionals based on local density (LDA) or generalized gradient (GGA) approximations give a metallic ground state, in contradiction with the Mott insulator nature of the compound. Some attempts to provide a realistic description of the strong correlations between the 5$f$-electrons of UO$_2$ at 0 K have obtained partial success by incorporating the Hubbard-like correction via the DFT+U scheme~\cite{Dorado2009}. Adding an on-site repulsion term, itinerant 5$f$-electrons become localized and $f$-orbitals fully occupied or empty. However, as pointed out by Dorado et al.~\cite{Dorado2013}, the orbital anisotropy derived from such a scheme makes the self-consistent algorithm extremely sensitive to the initial conditions, which leads to numerous metastable states. 

Including a portion of exact exchange, hybrid functionals might be an alternative for studying actinide oxides, if not for the fact that they also suffer from the intrinsic problem of metastable states typical of $f$-electron systems \cite{Kudin2002,Jollet2009}. Depending on the level of theory, the exchange-correlation functional employed, and on the convergence to different $f$-orbital occupations, a wide spread of values for properties such as the band gap can be obtained (see Refs~\cite{Thompson2011,Dorado2013} for a survey of results). Nonetheless, with the flavor of the DFT+U implementation affecting the energy and the ordering of the metastable states, the less biased nature of hybrid functionals, together with a scheme able to explore different $f$-orbital occupancies might help to reduce the uncertainty of such variations.

Going beyond hybrid functional DFT, a combination of DMFT with DFT provides a higher level of theory that is particularly well suited to describing $f$-electron materials. In addition to a better description of correlated interactions in comparison to DFT+U, it does not break the spin and orbital symmetry, and a subsequent reduced occurrence of metastable solutions may be obtained~\cite{Amadon2012}.  Finally, by selectively localizing the number of $f$-electrons, SIC has demonstrated solvency in describing $f$-electron delocalization in U compounds by correcting the unphysical self-interaction energy~\cite{Petit2002}. However, this does not avoid the introduction of tuning parameters. 

Although none of these methods has provided a conclusive response to the electronic and magnetic ground state of UO$_2$, all assumptions have proven to be well grounded in principles of evidence provided by experimental results. In all cases, results demonstrated the competence of chemical models to reproduce physical features such as structural parameters or energy-dependent magnetic configurations. At the same time, though, essential contradictions were found; in particular the different ground states found by several authors with similar methods, or the absence in literature of a solid explanation of the possible Jahn-Teller distortion, all of which led us to conclude that the actual ground state of cubic UO$_2$ needs further investigation.

In this paper we provide a systematic exploration of the occupancy space for UO$_2$ using a full hybrid functional approach, which allows us to avoid uncertainties due to the value of U and type of DFT+U approximation. Our methodology consists of the use of the PBE0 functional~\cite{Adamo1999} and $f$-orbital occupancy biasing rather than a strict constraint.  This has been implemented in the highly accurate wavelet-based BigDFT code\cite{Ratcliff2020}, which has been efficiently parallelized to allow calculations on much larger supercells than have previously been accessible using hybrid functionals~\cite{Ratcliff2018}. As well as comparing with previously reported pure and mixed states, our calculation scheme allows us to identify metastable states which are comparatively \emph{lower} in energy, and observe disordered occupancy patterns.

This study is open-ended and its main goal is to point out a new direction for future work. Specifically, we suggest that it would be worthwhile to develop new algorithms to more comprehensively explore such states, rather than finding them ``accidentally'' following a large number of algorithmic iterations.  Such work would be essential for the investigation of potential long range periodicity (spin waves), and to determine whether a definitive ground state could be found, or if there is a set (containing potentially a large number) of similar states which are all of more or less identical energy.  

This paper is arranged as follows.  In Section~\ref{sec:methods}, we outline the computational methods we have employed, in particular our approach to exploring metastable states.   In Section~\ref{sec:prelim}, we perform an initial systematic exploration of the occupancy space using hybrid functional DFT, while also calibrating the effects of supercell size and basis set convergence.  In Section~\ref{sec:landscape} we then screen a large number of potential $f$-electron occupancy matrices to build up a picture of the rich energy landscape.  Finally we conclude in Section~\ref{sec:conclusion}.

\section{Computational Methods}\label{sec:methods}

\subsection{DFT}

Due to  the high computational cost of hybrid functional calculations, the number of studies on UO$_2$ using DFT+U far exceeds those using hybrid functionals.  Furthermore, previous investigations of the ground state of UO$_2$ using hybrid functionals~\cite{Kudin2002,Prodan2006,Roy2008,Novak2006,Jollet2009} have been limited to small supercells, while the only detailed exploration of the occupancy space relied on the use of the ``exact exchange for correlated electrons'' (EECE) approximation to PBE0~\cite{Jollet2009}.
Such an approach resulted in a significantly lower band gap than for previous calculations with full PBE0 -- 2.0~eV for the lowest energy state found with EECE \emph{vs.}\ 3.1~eV for full PBE0~\cite{Prodan2006}, however this discrepancy might also be due to other differences in computational parameters, including the employed basis set, use of effective core potentials \emph{vs.}\ the projector augmented wave (PAW) approach~\cite{Blochl1994}, as well as differences in $k$-point sampling and initial guess.

In order to overcome the limitations to small cells, we make use of a recently implemented highly efficient parallelization of the exact exchange calculation in a wavelet basis set~\cite{Ratcliff2018}. This allows hybrid functionals calculations of large systems containing up to several hundred atoms at reasonable computational cost, without imposing additional approximations. Here PBE0 has been used for all calculations. Importantly, this approach employs a systematic basis set and thus does not require any (uncontrolled) approximations in either the basis set or the calculation of the exact exchange. Indeed, the reduction in computational cost comes from an efficient parallelization scheme, rather than any inherent approximations.  We use this approach as implemented in the BigDFT code, which is optimally adapted to scale on leadership-class computers.

DFT calculations were performed in cubic supercells of UO$_2$, for which the 96 atom ($2 \times 2 \times 2$) structure is depicted in Fig.~\ref{fig:uo2_struc}.  No relaxation of the atomic positions was performed.
Following the general consensus in the literature, we used the 1-$\mathbf{k}$ AFM structure and neglected spin-orbit coupling, which has been shown to have only a small effect~\cite{Roy2008}.
The employed supercells were of three different sizes: 12 atoms ($1 \times 1 \times 1$),  96 atom ($2 \times 2 \times 2$ and 324 atoms ($3 \times 3 \times 3$).  In all cases, no $k$-point sampling was applied, with calculations at the $\Gamma$-point only.  Although this is a severe approximation for the smallest cells, the qualitative behaviour of the different metastable states was shown to be relatively insensitive to the system size, so that such a protocol was sufficient for an initial low accuracy screening, as will be shown in the following.    
We used pseudopotentials of the HGH form in the Krack variant~\cite{Krack2005}, with 14 and 6 electrons for U and O respectively.

In agreement with previous studies~\cite{Dorado2009,Jollet2009}, we observed that both the absolute energies and the energy ordering of various metastable states was affected when symmetry was imposed, and so no symmetry constraints were applied.  In order to improve convergence a number of unoccupied states (35\% of the number of occupied states) were also optimized during the calculation, which had the additional benefit of providing direct access to the band gap and the low energy unoccupied density of states (DoS).  In addition, a small finite temperature of 0.1~eV was used to aid convergence.

It was observed that frequent re-diagonalization of the Hamiltonian, rather than pure direct minimization, was necessary to ensure correct self-consistent field (SCF) convergence.  We therefore performed a diagonalization every 12 SCF iterations.  For all calculations convergence was reached when the gradient fell below $10^{-4}$.  For the 12 atom screening calculations this was the sole convergence criterion, while for the higher quality calculations in larger cells it was further verified that the energy difference between two successive diagonalizations was less than 0.1~meV/atom.

\subsubsection{Occupancy Biasing}

In order to avoid convergence to a metastable state, various strategies have been proposed --  we refer to Ref.~\cite{Dorado2013} for further details on both the origins of metastable states in UO$_2$ and a summary of these schemes.
The most reliable of the suggested methods has proven to be the occupation matrix control (OMC) scheme~\cite{Amadon2008,Jomard2008,Dorado2009}.  In this approach, possible occupancy combinations are systematically explored and compared by imposing a given occupation matrix (OM) at the start of a calculation.  After sufficient iterations to ensure the calculation will remain trapped in the corresponding local minimum, the constraint is relaxed and the calculation is allowed to proceed to convergence.  The resulting energies for different occupancies are compared, and the OM corresponding to the lowest energy state (presumed to be the ground state) is thereafter imposed for subsequent calculations, such as those incorporating a defect.  Another popular approach is the U-ramping method~\cite{Meredig2010}, which relies on a gradual increase of the U parameter of DFT+U.  This approach has had some success, however has been shown to give higher energies than the OMC method~\cite{Meredig2010}.

The OMC approach has recently been exploited by Krack~\cite{Krack2015} to explore various different combinations of orbital occupancies for the two $f$-electrons using DFT+U, however the identified ground state does not agree with earlier DFT+U results from Dorado \emph{et al.}~\cite{Dorado2009,Dorado2010}, or with the ground state found using the EECE approach~\cite{Jollet2009}.  The use of such a scheme within a full hybrid functional approach could therefore be highly useful in clarifying the nature of the ground state of UO$_2$.

We therefore use an occupancy biasing scheme~\cite{Ratcliff2018}, wherein we introduce a bias towards a particular input OM, which is defined via the pseudopotential projectors.  This approach follows the same spirit as the OMC but is less strict in imposing the constraint.  We choose to apply this bias for the first 12 SCF iterations, after which the occupancy is free to evolve without constraints.

\begin{figure}
\centering
\begin{minipage}[t]{1.0\linewidth}
\centering
\includegraphics[scale=0.55]{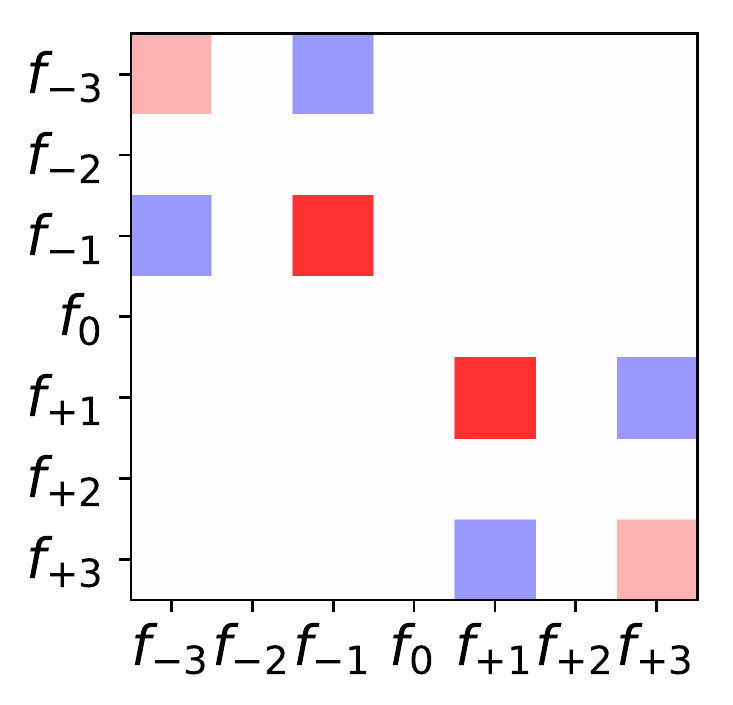}
\vspace{0pt}
\end{minipage}

\begin{minipage}[b]{1.0\linewidth}
\vspace{5pt}
\small
\centering
$\begin{pmatrix}
0.3                   & \textcolor{gray}{0.0} & $-0.4$ & \textcolor{gray}{0.0} & \textcolor{gray}{0.0} & \textcolor{gray}{0.0} & \textcolor{gray}{0.0} \\[4pt]
\textcolor{gray}{0.0} & \textcolor{gray}{0.0} & \textcolor{gray}{0.0} & \textcolor{gray}{0.0} & \textcolor{gray}{0.0} & \textcolor{gray}{0.0} & \textcolor{gray}{0.0} \\[4pt]
$-0.4$                & \textcolor{gray}{0.0} & 0.8 & \textcolor{gray}{0.0} & \textcolor{gray}{0.0} & \textcolor{gray}{0.0} & \textcolor{gray}{0.0} \\[4pt]
\textcolor{gray}{0.0} & \textcolor{gray}{0.0} & \textcolor{gray}{0.0} & \textcolor{gray}{0.0} & \textcolor{gray}{0.0} & \textcolor{gray}{0.0} & \textcolor{gray}{0.0} \\[4pt]
\textcolor{gray}{0.0} & \textcolor{gray}{0.0} & \textcolor{gray}{0.0} & \textcolor{gray}{0.0} & 0.8                   & \textcolor{gray}{0.0} & $-0.4$ \\[4pt]
\textcolor{gray}{0.0} & \textcolor{gray}{0.0} & \textcolor{gray}{0.0} & \textcolor{gray}{0.0} & \textcolor{gray}{0.0} & \textcolor{gray}{0.0} & \textcolor{gray}{0.0} \\[4pt]
\textcolor{gray}{0.0} & \textcolor{gray}{0.0} & \textcolor{gray}{0.0} & \textcolor{gray}{0.0} & $-0.4$                & \textcolor{gray}{0.0} & 0.3
\end{pmatrix}$
 \vspace{0pt}
\end{minipage}

\caption{Example of a non-diagonal occupancy matrix, $O$,  with mixing between $f$-orbitals, wherein four different orbitals have non-zero occupation. Top: color-resolved graphical representation of the OM, where dark (light) red (blue) squares indicate a large (small) positive (negative) value. Bottom: the OM itself.}
\label{fig:dmeg}
\end{figure}

The OM, $O$, of each atom may be specified independently.  Mixing between orbitals is allowed and included in the matrix representation in the form of non-diagonal terms. 
An example of an OM representing four $f$-orbitals with non-zero occupation and orbital mixing is shown in Fig.~\ref{fig:dmeg}. An equivalent graphical representation, which will be used in the following, is also depicted.
Since the occupancy constraint is not strictly imposed, there is no guarantee that the OM associated with a given atom remains normalised at the end of a calculation, however for ease of comparison the matrices are normalised before plotting.

\subsection{DFT+DMFT}

While both hybrid functional and DFT+U methods can suffer from convergence to different metastable states, DFT+DMFT can alleviate this problem by allowing the dynamical fluctuation of various many-body configurations of the $f$-electrons and treating the strong local correlation of those variants in an exact manner. Here, we use the DFT+DMFT method to compute the energetics, the OM, and the DoS of UO$_2$. In contrast to the DFT calculations, DMFT is performed using the 1-$\mathbf{k}$ AFM structure, with 2 U atoms per unit cell, and a lattice constant of 5.47~\AA.

The DFT+DMFT calculation is performed as follows. First, the DFT part is calculated using the projector augmented wave (PAW) method as implemented in the VASP code~\cite{Kresse1996a,Kresse1996b}.
The convergence of DFT is achieved using a plane-wave energy cut-off of 600~eV and a $k$-point mesh of 8$\times$8$\times$6. 
Once the DFT calculation is converged, the local subspace for the DMFT calculation is constructed using the maximally localized Wannier functions~\cite{Marzari2012} of U~$f$ and O~$p$ orbitals from the DFT non-spin-polarized band structure within the energy window of 11~eV. Then, the DMFT impurity problem for the U~$f$ orbital is solved using the continuous time quantum Monte Carlo method~\cite{Haule2015,Gull2011} including the hybridization of the correlated $f$~orbital with all other Wannier orbitals in the energy window. The DMFT self-consistency for the subspace has been performed for at least 40 iterations with a fine $k$-point mesh of 20$\times$20$\times$20. The Hubbard interaction $U=6$~eV has been used in previous DMFT calculations~\cite{Yin2011,Huang2017} and the Hund's coupling is $J=0.8$~eV. The so-called nominal form of the double counting potential is used~\cite{Haule2015}. More details of this DFT+DMFT calculation procedure can be found in Refs.~\cite{Park2014,Singh2021}.

\section{Benchmarking the Approach}\label{sec:prelim}

\subsection{Initial Screening}

For the purposes of narrowing down the space of configurations to investigate, in the first instance we have systematically explored the occupancy space wherein both up and down U atoms have the same input $f$-orbital occupancy, i.e.\ $f_{i(j)}^{\mathrm{UP}} \equiv f_{i(j)}^{\mathrm{DOWN}}$, with diagonal elements to the OM only, i.e.\ no imposed mixing between different $f$-orbitals.  This results in 21 possible input configurations.  However, once the occupancy biasing is switched off, there is no guarantee that the OM remains diagonal, or even identical for each U atom.  
 
Initial screening of the 21 combinations of $f$-occupancies was performed using a moderate grid spacing of $0.23$~\AA.
The input geometry and lattice constant of 5.42~\AA\ were taken from the Materials Project~\cite{Jain2013,MaterialsProject}.
Calculations were initially performed in a $2\times 2\times 2$ (96 atom) supercell.  The output OMs from select states were then used as inputs for calculations in $1\times 1\times 1$ (12 atom) and $3\times 3\times 3$ (324 atom) (super)cells to give an estimate of the error induced due to the lack of $k$-point sampling.
The relative energies, $\Delta E$, band gaps, E$_{\mathrm{g}}$, and final OMs are shown in Table~\ref{tab:energies}.   As can be seen, the size of the supercell has a significant effect on the band gap.  For the relative energies, there is a significant difference between the 12 atom values and those in larger cells, beyond which point the effect of increasing the supercell further is small, of the order of a few meV/atom.  Nonetheless, even the unit cell calculations are qualitatively similar, with the same states having the lowest energy for all setups.  Despite the lack of $k$-point sampling, one might therefore hope to use even the smallest cell for an initial approximate screening of different metastable states.

\begin{table*}
\centering
\begin{threeparttable}
\begin{tabular*} {1.0\textwidth}{l @{\extracolsep{\fill}} rrc c cc c cc c cc cc cc c cc}

\toprule
 &&& && \multicolumn{8}{c}{{PBE0}} && \multicolumn{5}{c}{DFT+U}  \\

 & &  &&   &&&     &&  &&&&& \multicolumn{2}{c}{U$_{\mathrm{eff}}$=3.96} && \multicolumn{2}{c}{U$_{\mathrm{eff}}$=2.00}\\

\cmidrule{6-13}  \cmidrule{15-19} \\[-2.5ex]

 & \multicolumn{2}{c}{Input} & Final  && \multicolumn{2}{c}{{$1\times 1\times 1$}}   &&  \multicolumn{2}{c}{{$2\times 2\times 2$}}   && \multicolumn{2}{c}{{$3\times 3\times 3$}} && \multicolumn{5}{c}{$4\times 4\times 4$}\\

& $f_i$& $f_j$ & OM &&  $\Delta E$ &  E$_{\mathrm{g}}$ &&  $ \Delta E$ & E$_{\mathrm{g}}$ && $\Delta E$ &  E$_{\mathrm{gap}}$ && $\Delta E$ &  E$_{\mathrm{g}}$ && $\Delta E$ &  E$_{\mathrm{g}}$\\

\cmidrule{1-4}   \cmidrule{6-7} \cmidrule{9-10} \cmidrule{12-13}  \cmidrule{15-16} \cmidrule{18-19} \\[-2.5ex]
- & -3 &  -1
& \begin{minipage}{.055\textwidth}
      \includegraphics[width=\linewidth]{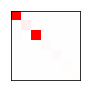}
    \end{minipage}
  && - & - && \multirow{2}{*}{\vspace{-15pt} 0.000} & \multirow{2}{*}{\vspace{-15pt}2.6} && - & -   && \multirow{2}{*}{\vspace{-15pt} 0.000} & \multirow{2}{*}{\vspace{-15pt}3.0} && \multirow{2}{*}{\vspace{-15pt}0.000} & \multirow{2}{*}{\vspace{-15pt}1.9} \\
A & +1 & +3
&\begin{minipage}{.055\textwidth}
      \includegraphics[width=\linewidth]{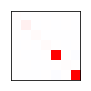}
    \end{minipage}
 && 0.000 & 2.3 && & && 0.000 & 3.0 \\
 
\cmidrule{1-4}   \cmidrule{6-7} \cmidrule{9-10} \cmidrule{12-13}  \cmidrule{15-16} \cmidrule{18-19} \\[-2.5ex]
B & 0 & +2 
& \begin{minipage}{.055\textwidth}
      \includegraphics[width=\linewidth]{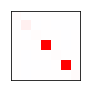}
    \end{minipage}
&& 0.007 & 2.5 && 0.003 &  2.9 &&  0.004 & 3.1 &&
0.003 &  3.4 &&  0.004 &  2.2 \\

\cmidrule{1-4}   \cmidrule{6-7} \cmidrule{9-10} \cmidrule{12-13}  \cmidrule{15-16} \cmidrule{18-19} \\[-2.5ex]
- & -3 &  +1 
&\begin{minipage}{.055\textwidth}
      \includegraphics[width=\linewidth]{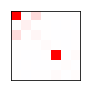}
    \end{minipage}
&& - & - && \multirow{2}{*}{\vspace{-15pt} 0.003} & \multirow{2}{*}{\vspace{-15pt} 2.9}  && - & - &&
 \multirow{2}{*}{\vspace{-15pt} 0.009} & \multirow{2}{*}{\vspace{-15pt} 3.3} && \multirow{2}{*}{\vspace{-15pt} 0.009} & \multirow{2}{*}{\vspace{-15pt} 2.3} \\
C & -1 &  +3 
& \begin{minipage}{.055\textwidth}
      \includegraphics[width=\linewidth]{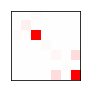}
    \end{minipage}
&& -0.001 & 2.5 && & && 0.005 & 3.3 \\

\cmidrule{1-4}   \cmidrule{6-7} \cmidrule{9-10} \cmidrule{12-13}  \cmidrule{15-16} \cmidrule{18-19} \\[-2.5ex]
- & -2 &  0 
&\begin{minipage}{.055\textwidth}
      \includegraphics[width=\linewidth]{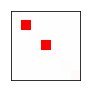}
    \end{minipage}
&& - & - &&  0.304 &  1.8 &&  { -} & { -} &&
 { -} & { -} && { -} & { -}\\
 
\cmidrule{1-4}   \cmidrule{6-7} \cmidrule{9-10} \cmidrule{12-13}  \cmidrule{15-16} \cmidrule{18-19} \\[-2.5ex]
- & -2 &  +2 
&\begin{minipage}{.055\textwidth}
      \includegraphics[width=\linewidth]{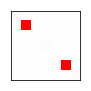}
    \end{minipage}
&& - & - &&  0.423 &  1.1 &&  { -} & { -} &&
 { -} & { -} && { -} & { -} \\ 

\midrule \\[-3.5ex]
\midrule \\[-2.5ex]
    
\multirow{2}{*}{\vspace{-8pt} D} 
& -3 &  +3& \multirow{2}{*}{\begin{minipage}{.055\textwidth}
      \includegraphics[width=\linewidth]{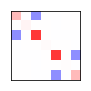}
    \end{minipage}}\vspace{8pt}
 && \multirow{2}{*}{\vspace{-8pt} -0.047} & \multirow{2}{*}{\vspace{-8pt} 2.6} &&\multirow{2}{*}{\vspace{-8pt} -0.025} & \multirow{2}{*}{\vspace{-8pt}  3.4}  && \multirow{2}{*}{\vspace{-8pt} -0.024} & \multirow{2}{*}{\vspace{-8pt} 3.6} && 
 { -} & { -} && { -} & { -} \\
& -1 &  +1 & &&&  & &&   & &&&&
 0.054 &  3.1 &&  0.054 &  2.0   \\
\cmidrule{1-4}   \cmidrule{6-7} \cmidrule{9-10} \cmidrule{12-13}  \cmidrule{15-16} \cmidrule{18-19} \\[-2.5ex]

\multirow{2}{*}{\vspace{-8pt} E} & -3 &  +2
& \multirow{2}{*}{\begin{minipage}{.055\textwidth}
      \includegraphics[width=\linewidth]{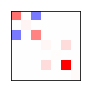}
    \end{minipage}}\vspace{8pt}
 && \multirow{2}{*}{\vspace{-8pt} -0.049} & \multirow{2}{*}{\vspace{-8pt} 2.5} && \multirow{4}{*}{\vspace{-15pt} -0.025} & \multirow{4}{*}{\vspace{-15pt} 3.1} && \multirow{2}{*}{\vspace{-8pt} -0.023} & \multirow{2}{*}{\vspace{-8pt}3.4} &&
 \multirow{4}{*}{\vspace{-15pt}  -} & \multirow{4}{*}{\vspace{-15pt}  -} && \multirow{4}{*}{\vspace{-15pt}  -} & \multirow{4}{*}{\vspace{-15pt}  -} \\
& -1 &  +2 &&&& \\
\multirow{2}{*}{\vspace{-8pt} -} &  +1 &  +2 
& \multirow{2}{*}{\begin{minipage}{.055\textwidth}
      \includegraphics[width=\linewidth]{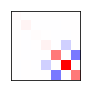}
    \end{minipage}}\vspace{8pt}
&& \multirow{2}{*}{\vspace{-8pt} -} & \multirow{2}{*}{\vspace{-8pt} -} & && && \multirow{2}{*}{\vspace{-8pt} -} & \multirow{2}{*}{\vspace{-8pt} -}\\
& +2 &  +3& &&& \\

\cmidrule{1-4}   \cmidrule{6-7} \cmidrule{9-10} \cmidrule{12-13}  \cmidrule{15-16} \cmidrule{18-19} \\[-2.5ex]
F & -3 &  0
& \begin{minipage}{.055\textwidth}
      \includegraphics[width=\linewidth]{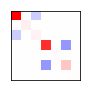}
    \end{minipage}
 && -0.010 & 2.6 && \multirow{2}{*}{\vspace{-15pt} 0.001} & \multirow{2}{*}{\vspace{-15pt} 3.2} && 0.003 & 3.5&&
 \multirow{2}{*}{\vspace{-15pt}$<$0.001} & \multirow{2}{*}{\vspace{-15pt} 3.5} && \multirow{2}{*}{\vspace{-15pt} 0.005} & \multirow{2}{*}{\vspace{-15pt} 2.2} \\
- & 0 &  +3 
& \begin{minipage}{.055\textwidth}
      \includegraphics[width=\linewidth]{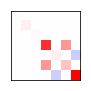}
    \end{minipage}
&& - & - & && && - & - \\

\cmidrule{1-4}   \cmidrule{6-7} \cmidrule{9-10} \cmidrule{12-13}  \cmidrule{15-16} \cmidrule{18-19} \\[-2.5ex]
G & -1 &  0 
& \begin{minipage}{.055\textwidth}
      \includegraphics[width=\linewidth]{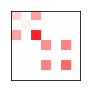}
    \end{minipage}
&& -0.004  & 2.6 && \multirow{2}{*}{\vspace{-15pt} 0.001} & \multirow{2}{*}{\vspace{-15pt} 2.8} && 0.003 & 3.2 &&
 \multirow{2}{*}{\vspace{-15pt} -} & \multirow{2}{*}{\vspace{-15pt} -} && \multirow{2}{*}{\vspace{-15pt} -} & \multirow{2}{*}{\vspace{-15pt} -} \\
- & 0 &  +1 
&\begin{minipage}{.055\textwidth}
      \includegraphics[width=\linewidth]{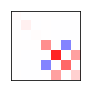}
    \end{minipage}
&& - & - &&&&& - & - \\
 
\midrule \\[-3.5ex]
\midrule \\[-2.5ex]
    
H & -3 &  -2 
& \begin{minipage}{.055\textwidth}
      \includegraphics[width=\linewidth]{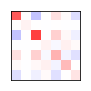}
    \end{minipage}
&& - & - && { -0.026} & { 3.2}  && \multirow{2}{*}{\vspace{-15pt} -} & \multirow{2}{*}{\vspace{-15pt} -} &&
 \multirow{2}{*}{\vspace{-15pt} -} & \multirow{2}{*}{\vspace{-15pt} -} &&
 \multirow{2}{*}{\vspace{-15pt} -} & \multirow{2}{*}{\vspace{-15pt} -}\\
- & -2 &  +3 
& \begin{minipage}{.055\textwidth}
      \includegraphics[width=\linewidth]{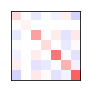}
    \end{minipage}
&& - & - && { -0.019} & { 3.3} \\

\cmidrule{1-4}   \cmidrule{6-7} \cmidrule{9-10} \cmidrule{12-13}  \cmidrule{15-16} \cmidrule{18-19} \\[-2.5ex]
- & -2 &  -1 
& \begin{minipage}{.055\textwidth}
      \includegraphics[width=\linewidth]{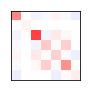}
    \end{minipage}
&& - & - &&  { -0.018} & { 3.1}  && \multirow{2}{*}{\vspace{-15pt} -} & \multirow{2}{*}{\vspace{-15pt} -} &&
 \multirow{2}{*}{\vspace{-15pt} -} & \multirow{2}{*}{\vspace{-15pt} -} &&
 \multirow{2}{*}{\vspace{-15pt} -} & \multirow{2}{*}{\vspace{-15pt} -} \\
- & -2 &  +1 
& \begin{minipage}{.055\textwidth}
      \includegraphics[width=\linewidth]{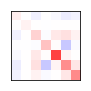}
    \end{minipage}
&& - & - && { -0.016} & { 3.1}\\

  \bottomrule
\end{tabular*}
\end{threeparttable}
\caption{Initial screening of metastable states using a (cubic) lattice constant of 5.42~\AA\ and a grid spacing of 0.23~\AA\ for different supercells.
$\Delta E$ is the relative energy (relative to state A) in eV/atom, while E$_{\mathrm{g}}$ is the band gap (in eV). Final OMs (averaged over all U atoms) are shown for different input occupancies ($f_i$ and $f_j$) for the $2 \times 2 \times 2$ calculations.
The results are grouped as pure, mixed and disordered states from top to bottom.  Where available, DFT+U results (taken from Ref.~\cite{Krack2015}) with the same \emph{input} occupancies are also shown.}  
\label{tab:energies}
\end{table*}

Although a diagonal OM was imposed for each calculation, a large number of the calculations eventually converged to a rather different state, presumably due to the lack of a strict imposition of the occupancy constraint.   To aid in the discussion, from now on we categorize the output occupancy matrices as either: `pure' states, which have (more or less) retained their diagonal input occupancies; `mixed' states, for which mixing with other states has occurred, as indicated by non-negligible  off-diagonal elements in the OM; and `disordered' or noisy states, wherein the final occupancies have strayed far from the input values.  In the latter case, different U atoms might also have quite different final OMs, while the calculations also took a very large number (of the order of 1000s) of SCF iterations to converge. 

For context, results for DFT+U (extracted from Krack~\cite{Krack2015}) are also shown where available.   Where OMs remained pure, the agreement between PBE0 and DFT+U for $U_{\mathrm{eff}}=3.96$ is good, with the same ordering of states and similar band gaps, although some disagreement is expected due to differing computational protocols (cell dimensions, supercell size etc.).
Although the mixed OMs we find differ in their exact form, the existence of low energy mixed states agrees with previous observations by Dorado \emph{et al.}~\cite{Dorado2009,Dorado2010}, who in contrast to Krack explicitly searched for such states.  However, the appearance of \emph{low energy} disordered states has not previously been observed, and is worthy of further investigation.

There is also a significant spread of band gaps, with the values in a similar range to previous results employing hybrid functionals~\cite{Kudin2002,Prodan2006,Roy2008,Novak2006,Jollet2009}, although at the higher end.  In all cases the gap is notably higher than the experimental value, suggesting that in future work it might be interesting to either consider different hybrid functionals and/or vary the fraction of exact exchange.
It should be noted that in each case a non-metallic state was reached, in contrast to previous calculations with DFT+U where metallic states were sometimes found~\cite{Dorado2009}.  Likewise, each case was able to eventually reach convergence, albeit sometimes at the cost of a very high number of SCF iterations.  Since the OM is less strictly imposed, it is presumably the case that small levels of noise exist in the OM.  In cases where the corresponding state is high in energy or even unstable, this noise is able to gradually accumulate, providing a pathway to a lower energy state which either adds some additional mixing or departs entirely from the input OM, as in the case of the disordered states.
In the following we focus on further exploring these low energy states.

\subsection{Basis Set Convergence}

While these results providing a good starting point, it is important to both converge the basis set and optimize the lattice constant in order to obtain more quantitative results, before further investigating the low relative energies of the mixed and disordered states.  In order to determine whether or not the different states are affected by the basis set size, we take a representative selection of states as indicated by labels A-H in Table~\ref{tab:energies}.

The relative energies of pure and higher energy mixed states are not strongly affected by the grid spacing i.e.\ they all converge similarly with respect to the basis.  The lowest energy mixed states, and to a lesser extent the disordered state H, converge rather differently, so that the relative energies change significantly with increasing basis size.  Nonetheless, qualitatively, the results of the initial screening hold true.  Interestingly, the disordered state becomes the lowest energy state by around 6~meV/atom compared to the next lowest state.
Unlike the relative energies, the band gaps are insensitive to the basis set size.
Detailed results can be found in the Supplementary Information.
Based on these observations, a grid spacing of $h=0.18$~\AA\ can be considered sufficient for high accuracy results, while 
$h=0.2$~\AA\ will be used for further screening calculations where only a moderate accuracy is required.

\subsection{Lattice Constant}

\begin{figure}
\centering
\includegraphics[scale=0.38]{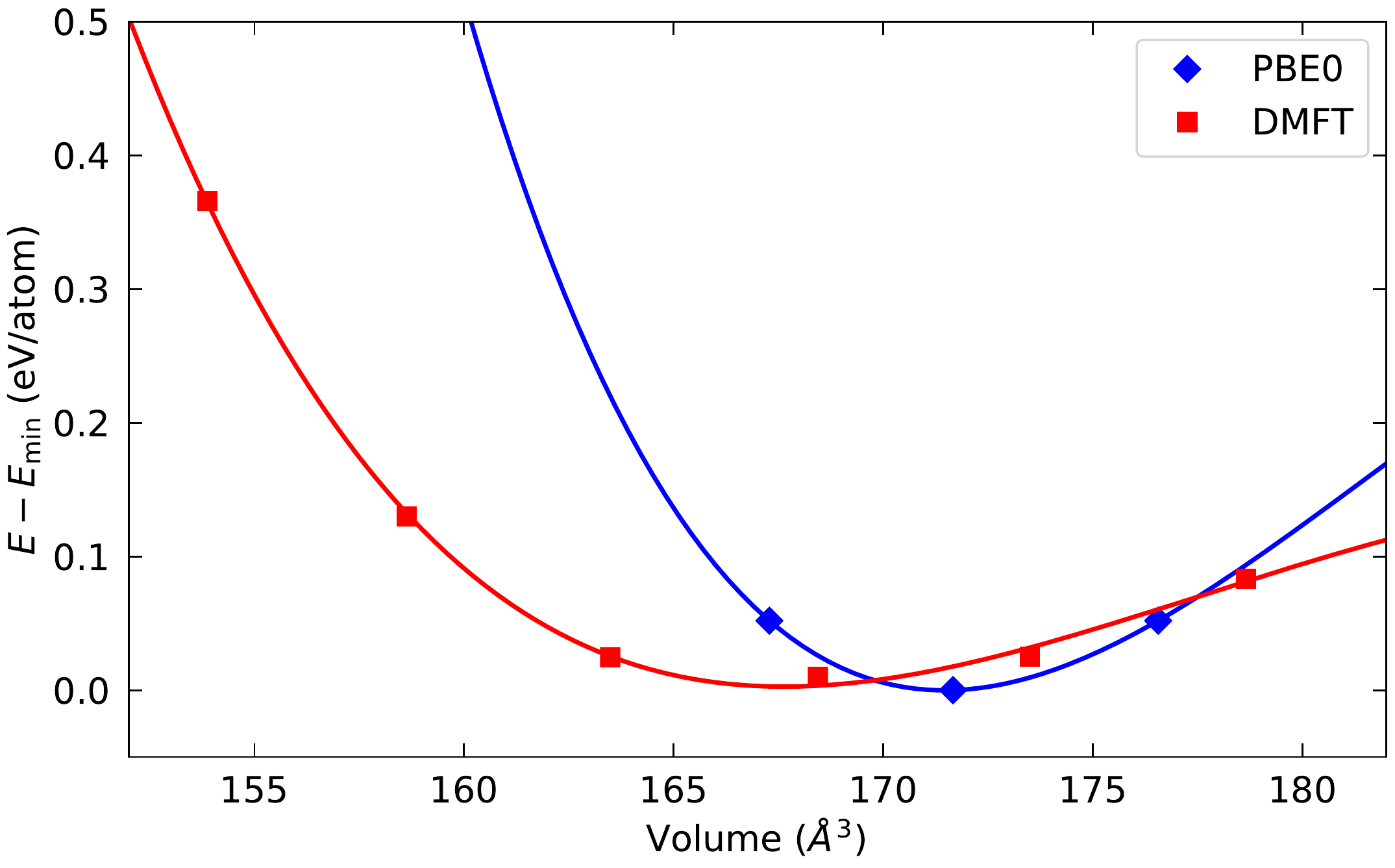}
\caption{Volume \emph{vs.}\ energy for state D in a $3\times 3 \times 3$ supercell with PBE0, compared to the DMFT results. Also shown is the fit to the Birch-Murnaghan equation of state. \label{fig:lattice}}
\end{figure}

Following from the above considerations, the lattice constant was also explored.  Taking the accurate wavelet grid spacing of $0.18$~\AA, the lattice constant was varied for states A-H using the supercell of 96 atoms, restricting to a cubic supercell.   All states were found to have an equilibrium lattice constant of 5.60~\AA, as shown in the Supplementary Information.  Since the lattice constant was found to be independent of the orbital occupancy, calculations were performed for a supercell of 324 atoms for state D only, as depicted Fig.~\ref{fig:lattice} alongside the DMFT results.  The results are also fitted to the third order Birch-Murnaghan equation of state~\cite{Birch1947}, giving equilibrium lattice constants of 5.56~\AA\ for PBE0, compared to 5.51~\AA\ for DMFT.  Due to the high computational cost of the large supercell calculations, only three lattice constants are explored, thus the DMFT value for the pressure derivative of the bulk modulus, $B_0'$, is employed for the PBE0 fit. This approach gives the same equilibrium lattice constant as fitting a quadratic polynomial.  Although the PBE0 result is at the high end of the range of results found in the literature, with the DMFT result also being closer to experiment, the relative energies of the different states are not strongly affected by varying the lattice constant within the range of values considered, and thus a small change would have little impact on the conclusions which are drawn in the following.  Therefore, all subsequent PBE0 calculations use a lattice constant of 5.56~\AA.

\section{Exploring the Energy Landscape}\label{sec:landscape}

\subsection{Single Occupancy Constraint}

\begin{figure}
\centering
\includegraphics[width=0.5\textwidth]{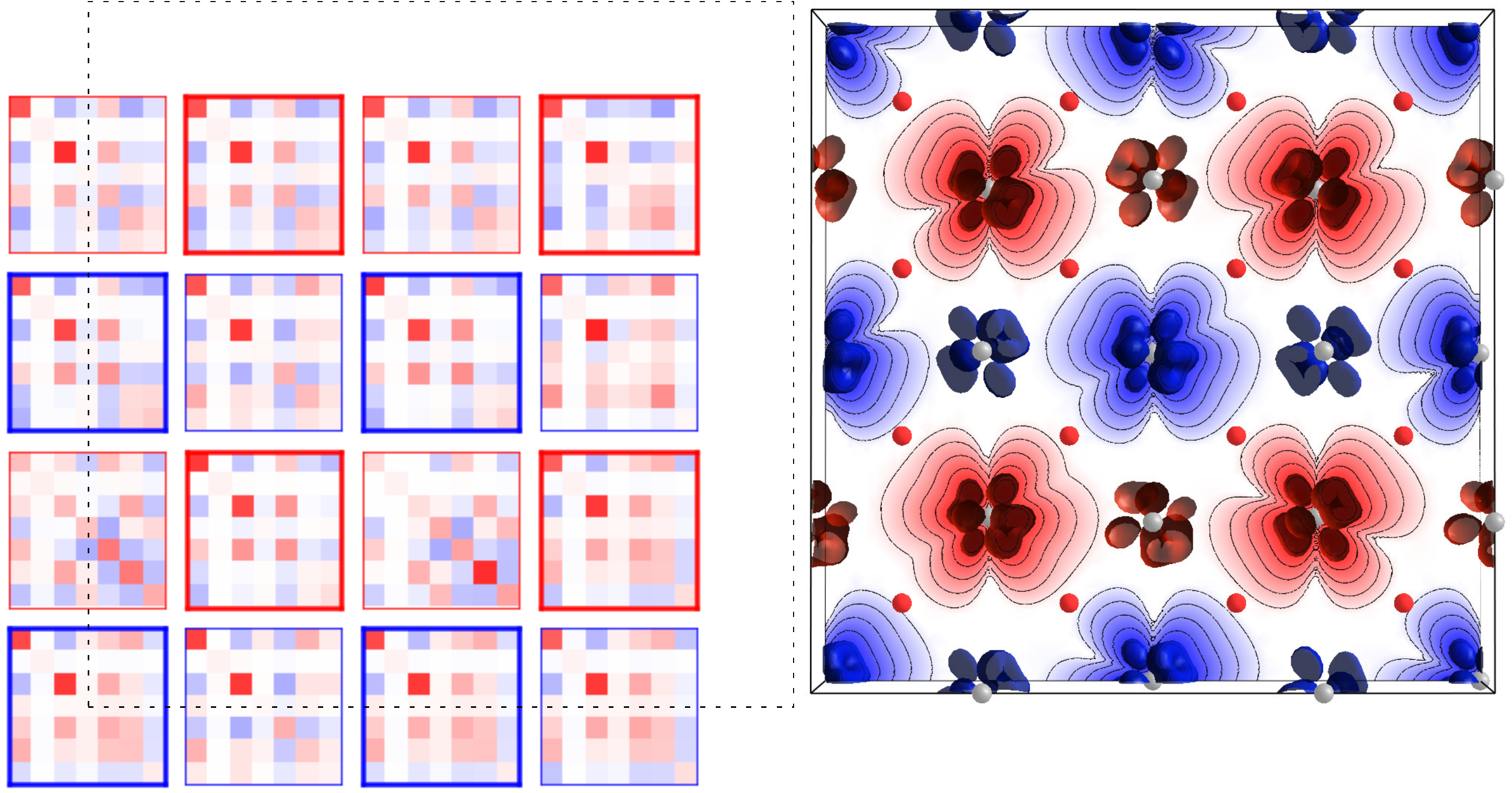}
\caption{Final OMs [left] and spin density [right] for a cross-section of the supercell for state H.  The type of U atom and position relative to the plane is indicated by the colour of the OM borders -- thick (thin) lines indicate an atom which is in (above) the plane, while blue (red) borders and correspondingly densities, indicate spin up (down).  The centre of each OM indicates the position of the U atom within the plane, while the dashed outline indicates the boundaries of the supercell. The depicted isosurfaces correspond to a density of $\pm 0.1$~$e a_0^{-3}$.
}\label{fig:disordered}
\end{figure}

So far, we have established the existence of a set of new low energy metastable states.  However, there are two possible reasons for the low energies.  First, the final OMs are far from pure, having significant off-diagonal components, which is potentially already sufficient to lower the energy.  Second, taking the case of state H for example, different atoms were found to have differing OMs, as can be seen in Fig.~\ref{fig:disordered}.  
This might also be reasonably expected to lower the energy -- for example it was previously shown using DFT+U that a lower energy might be found by relaxing the constraint that $f_{i(j)}^{\mathrm{UP}} \equiv f_{i(j)}^{\mathrm{DOWN}}$~\cite{Krack2015}.
In the first instance we consider the first point, i.e.\ we continue to investigate states wherein \emph{all} atoms within the supercell have the same input occupancy, i.e.\ $f_{i(j)}^{\mathrm{UP}} \equiv f_{i(j)}^{\mathrm{DOWN}}$, or more generally $O^{\mathrm{UP}} \equiv O^{\mathrm{DOWN}}$.

In order to investigate a large number of states, we take the smallest cell (12 atoms) with the aim of screening out states which have significantly higher energy.
Taking the final OMs from the above calculations, we already have many possible input OMs.  First, those for all pure and mixed states may be directly used.
Second, for the disordered states where different atoms might have different OMs, we take the complete set of final OMs for all atoms within each supercell, and then discard those OMs which are very similar.  To further expand the range of potential inputs we take also other input OMs from DFT+U calculations in the literature, notably Dorado's lowest energy off-diagonal OM~\cite{Dorado2009} and Thompson and Wolverton's minimum energy state, which was obtained using U-ramping~\cite{Thompson2011}.  In order to avoid repeating calculations, similar OMs were screened out at each stage by using the average absolute difference between two OMs as an indicator of distinctness.
Initial calculations were performed for each distinct OM in the above set using a moderate grid spacing ($h=0.2$~\AA). All OMs giving an energy less than $10$~meV/atom above state A were retained, resulting in 48 candidate OMs.

Although the current methodology was able to find new low energy states, the exorbitant computational cost and the lack of ability to systematically explore such states requires a new approach to more effectively investigate the existence of other low energy OMs.  Since it is presumed that the lowest energy states were found due to small amounts of noise opening up new search directions, we now also deliberately add randomness to the candidate OMs.  This is achieved in two ways: first by adding small amounts of random noise to a given OM, and second by changing the sign of a given off-diagonal element with a certain probability.  In both cases, symmetry was imposed on the OM by ensuring $O_{ij}\equiv O_{ji}$.  For each candidate OM, such a procedure was followed for ten steps, each time retaining distinct OMs with an energy less than state A.  This resulted in a larger set of OMs, comprising 178 in total.

\begin{figure*}
\centering
\subfigure[$1\times 1 \times 1$, $O^{\mathrm{UP}} \equiv O^{\mathrm{DOWN}}$] {\includegraphics[scale=0.37]{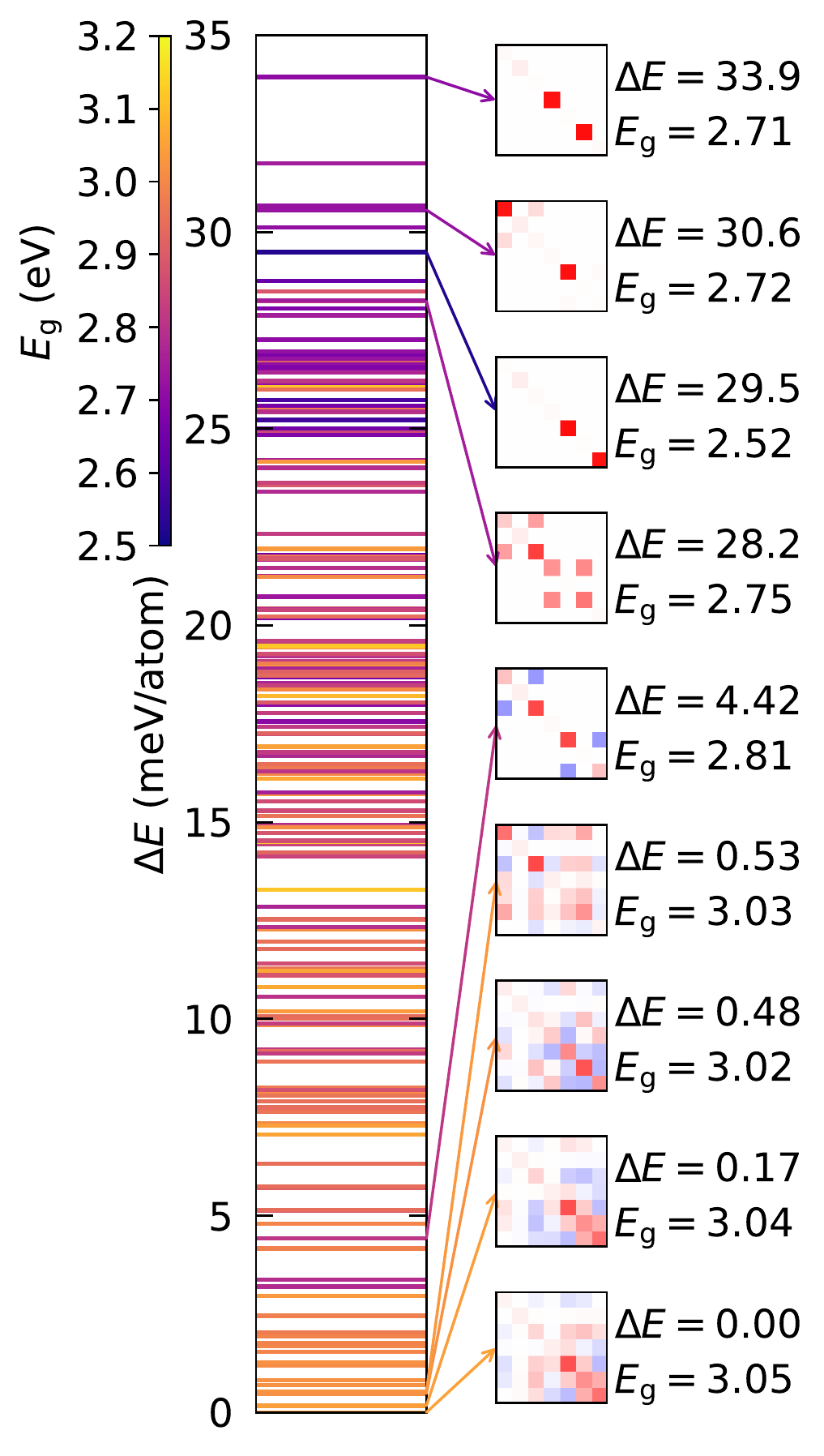}\label{fig:s111h}}
\subfigure[$1\times 1 \times 1$, $O^{\mathrm{UP}} \neq O^{\mathrm{DOWN}}$] {\includegraphics[scale=0.37]{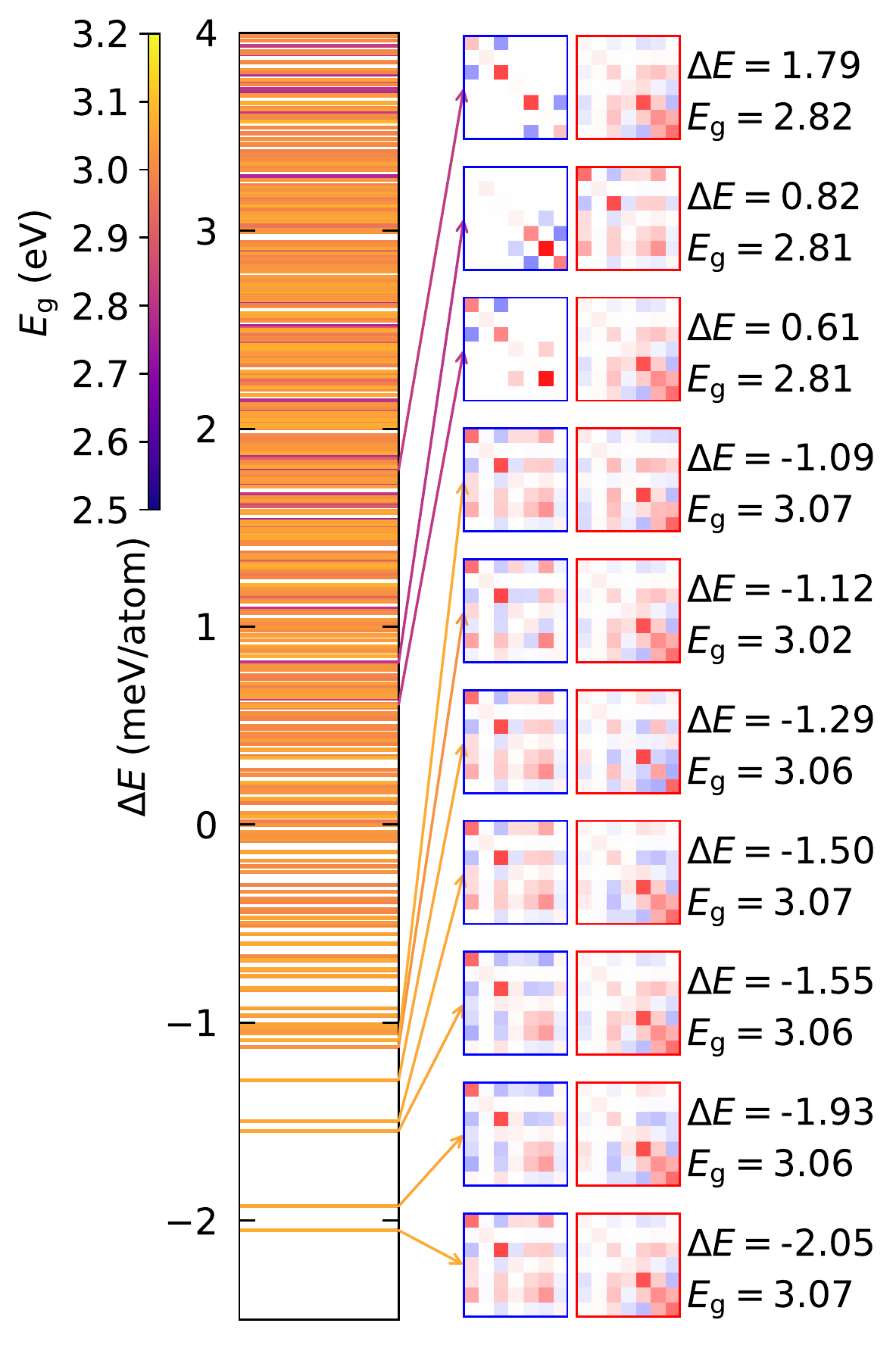}\label{fig:s111hm}}
\subfigure[$2\times 2 \times 2$, $O^{\mathrm{UP}} \equiv O^{\mathrm{DOWN}}$] {\includegraphics[scale=0.37]{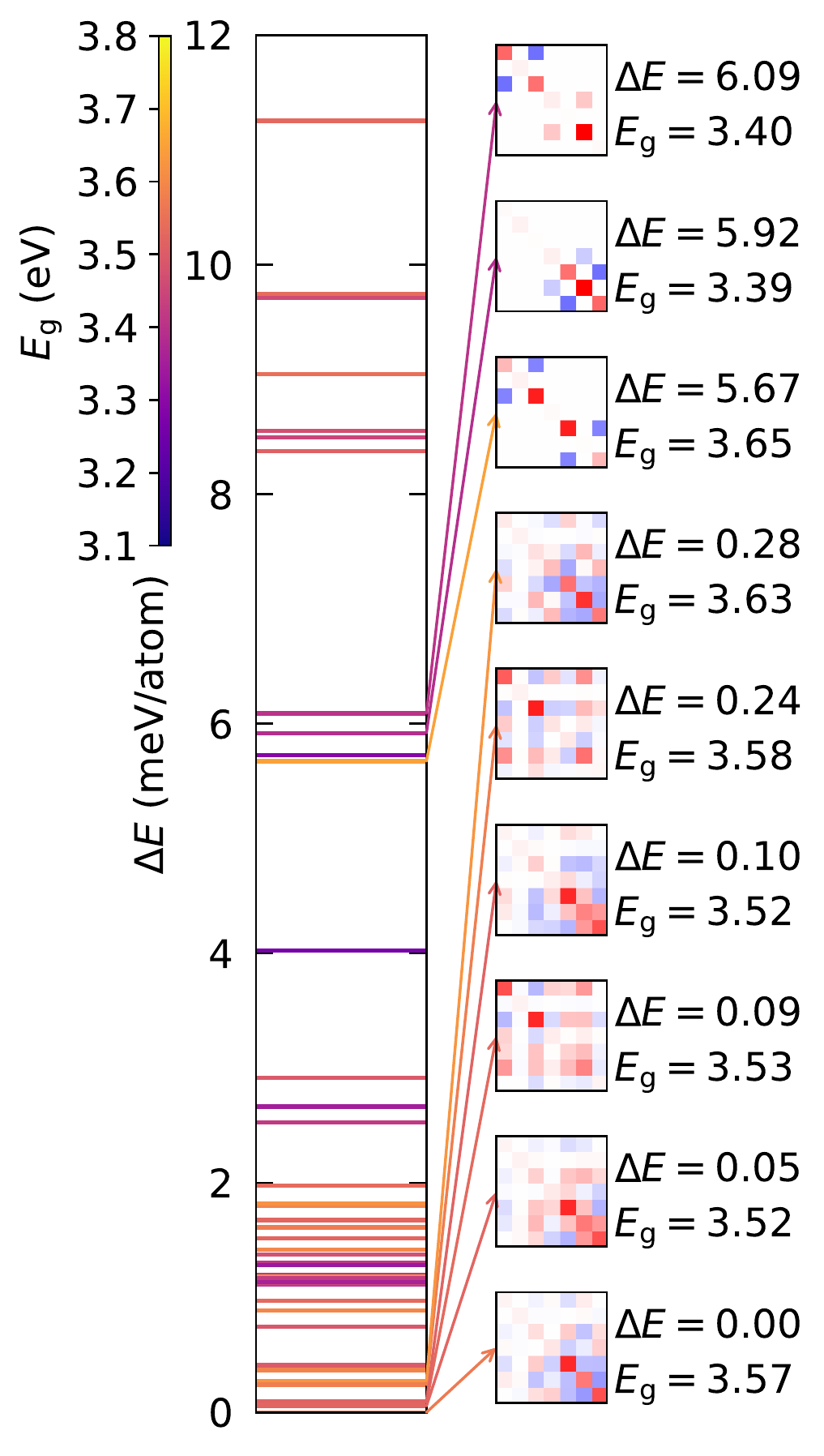}\label{fig:s222h}}
\subfigure[$2\times 2 \times 2$, $O^{\mathrm{UP}} \neq O^{\mathrm{DOWN}}$] {\includegraphics[scale=0.37]{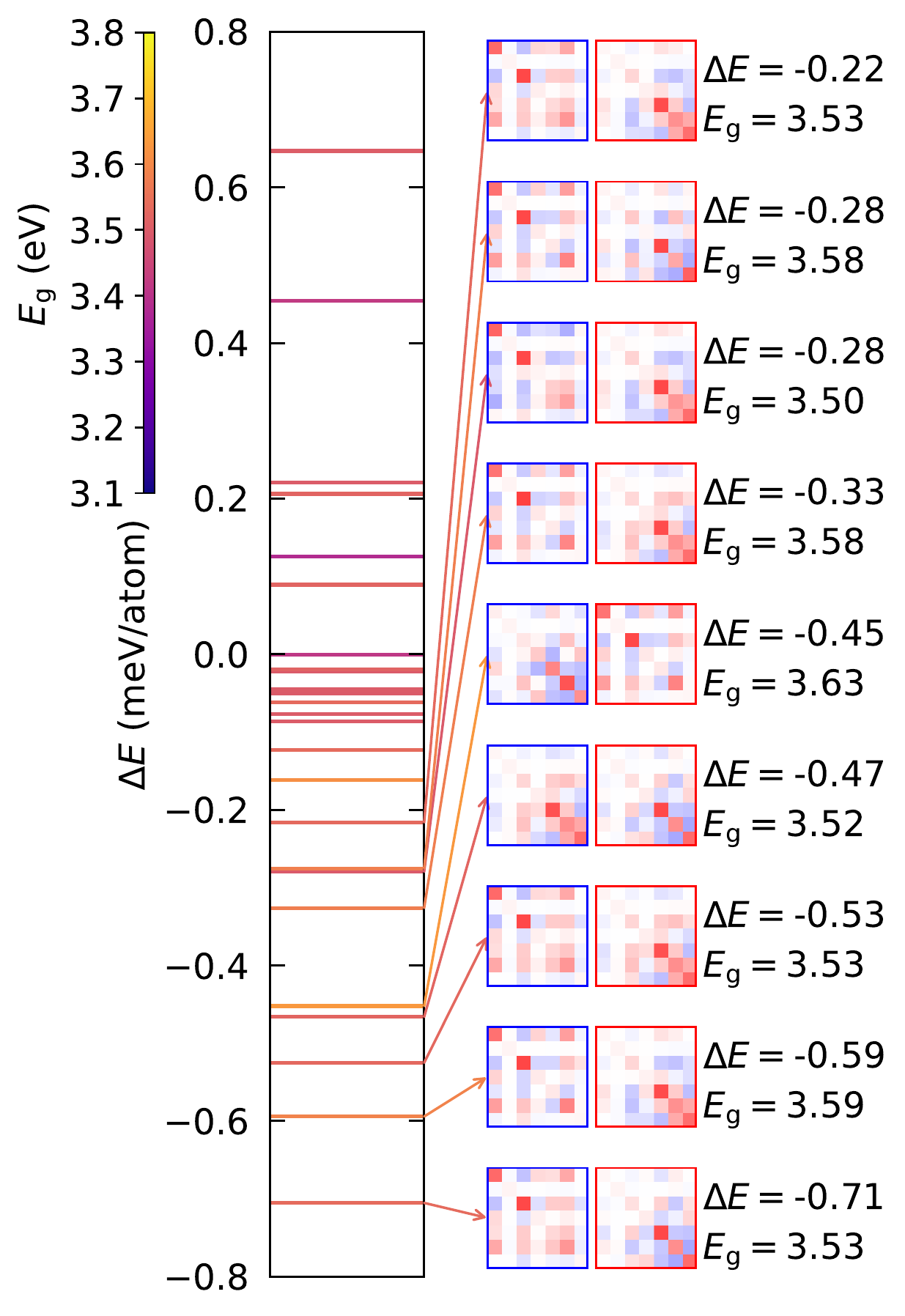}\label{fig:s222hm}}
\caption{Relative energies, $\Delta E$, and band gaps, $E_{\mathrm{g}}$, (as indicated by the line colour) for the range of explored states in different supercell sizes and restrictions on the OMs, as indicated.  The final OMs are depicted for select states, while for~\ref{fig:s111hm} and~\ref{fig:s222hm} the OMs for up (down) U atoms are depicted with a blue (red) border.
}\label{fig:uo2_s}
\end{figure*}

Each of these calculations was then repeated using a more accurate grid spacing ($h=0.18$~\AA).  While the basis set had some impact on the relative energies, this was relatively small, with only a few states giving significantly different results (see Supplementary Information).  Furthermore, a large enough energy range was considered to be confident that no low energy states would be incorrectly excluded from further consideration.  Finally, all states within 10~meV/atom of the new lowest energy state (44 OMs) were then calculated in a 96 atom cell, again with a grid spacing of $h=0.18$~\AA.  The differences in relative energies, and even more so the band gap, are more significant when the cell size is increased (see Supplementary Information).  Nonetheless, by calculating all states within 10~meV/atom of the lowest energy state for the smallest cell, we can be reasonably sure that at least those states which are within a few meV/atom of the minimum for the larger cell have been included.   The relative energies, band gaps and OMs of select states for the 12 and 96 atom calculations are depicted in Figs.~\ref{fig:s111h} and~\ref{fig:s222h} respectively.

Although such calculations might be considered merely qualitative rather than quantitative, the calculations in the 12 atom cell show a very rich energy landscape, with a large number of states which are lower in energy than state A, Krack's previously identified ground state (depicted in Fig.~\ref{fig:s111h} at around 30~meV/atom higher in energy than the new lowest energy state). 
Going to the larger unit cell, there are seven states within 1~meV/atom of the new lowest state (depicted on the left of Fig.~\ref{fig:s222h}, all of which have disordered OMs.  The lowest non-disordered states (for which the OMs are depicted in Fig.~\ref{fig:s222h}) are more than 5~meV/atom higher in energy than the lowest state.
Although not shown on Fig.~\ref{fig:s222h}, we also recalculated the energy of state H using the same computational parameters.
It is interesting to note that the new lowest energy state is lower in energy than state H, albeit by less than 1~meV/atom, despite having the additional constraint that all U atoms have the same OM.  One might therefore speculate that finding the optimal OM is at least as important in minimizing the energy as allowing the OM to vary between U atoms.

For the smallest cell there also appears to be a tendency for lower energy states to have a higher band gap, in line with previous observations~\cite{Jollet2009}.  As shown in the Supplementary Information there is indeed a negative correlation between the two.  However the error when calculating the band gap for the 12 atom cell is very high, so that such a conclusion is not reliable.  Indeed, as is also shown in the Supplementary Information, the correlation between the band gap in the two supercell sizes is weak.  In the 96 atom supercell, there remains a negative correlation between relative energy and band gap but it is noticeably weaker.  However, the energy range considered is much smaller so that further data points would be needed in order to draw a firm conclusion.

It is important to reiterate that although we have previously shown that the relative energies are not strongly affected by further increasing the supercell size, and despite the use of a relatively well converged systematic basis, the expected error in the relative energies is higher than the very small differences in energy between the lowest energy states.  In other terms, even though the (numerical) ``noise'' associated with the computational setup is small, the signal that we are seeking is even smaller.  Therefore, it is not possible to conclude that a given state is indeed the ground state of UO$_2$.  Furthermore, it is entirely possible that by further varying the OMs one could find a large number of other metastable states with similar energies.  Nonetheless, it is clear that the energy landscape is inherently more complex than has previously been shown.  It is also evident that having a disordered OM can result in significantly lower energy states, even when all U atoms have the same imposed OM.

We also observe that although the band gap is not yet converged with respect to system size, it nonetheless varies by around 0.4~eV within the energy range considered.  To be more quantitative such calculations should be repeated in a larger supercell, however given the large number of OMs under consideration, such simulations would be very costly.  Nonetheless such a figure serves to illustrate the complexity in the energy landscape of UO$_2$, wherein even states with very close energies can have very different band gaps.

\subsection{Combining Occupancy Constraints}

Having considered in detail the case where all atoms are constrained to have the same occupancy, we now relax the constraint, by imposing different occupancies on U atoms associated with up and down spins, i.e.\  $O^{\mathrm{UP}} \neq O^{\mathrm{DOWN}}$.  In the more general case, this could extend to further allowing \emph{all} U atoms in a given supercell to have different occupancies. However, given the vast number of possible resulting combinations, such simulations would be extremely costly, and such considerations are therefore left for future work.

We take again the 44 OMs which are less than 10~meV/atom above the lowest energy state in the 12 atom cell, i.e.\ those which were also calculated in the 96 atom cell.  We then consider every pair $\{i,j\}$ of such OMs, where $i<j$, and perform calculations for $\{O_{i}^{\mathrm{UP}}$, $O_{j}^{\mathrm{DOWN}}\}$ in the 12 atom cell with $h=0.2$~\AA.  
All states less than 4 meV/atom above the previously identified lowest state in the same computational setup were kept, giving 460 OM setups.  These calculations were then repeated for $h=0.18$~\AA, the results for which are depicted in Fig.~\ref{fig:s111hm}. All calculations within 1.5 meV/atom of new lowest state were repeated in 96 atom cell, giving 26 OM setups, for which the results are shown in Fig.~\ref{fig:s222hm}.
As above, such a procedure does not guarantee that every low energy state has been included in the set of 96 atom calculations, however it represents a compromise wherein the majority of low energy states are expected to have been found without having to calculate a large number of high energy states.
Indeed, of the 26 setups which were considered, 19 have a lower energy than the previously identified lowest state where $O^{\mathrm{UP}} \equiv O^{\mathrm{DOWN}}$, with the lowest being 0.7~meV/atom lower in energy.  For each setup in the energy range considered, both the up and down OMs were disordered.  Given the small difference of less than 1~meV/atom between the lowest energy setup with $O^{\mathrm{UP}} \neq O^{\mathrm{DOWN}}$ compared to $O^{\mathrm{UP}} \equiv O^{\mathrm{DOWN}}$, it is clear that allowing a disordered OM has a much a stronger effect than allowing different atoms to have different occupancies. However, as stated it is possible that the energy could be further reduced by allowing all U atoms to have different occupancies.

\begin{figure*}
\centering
\includegraphics[scale=0.39]{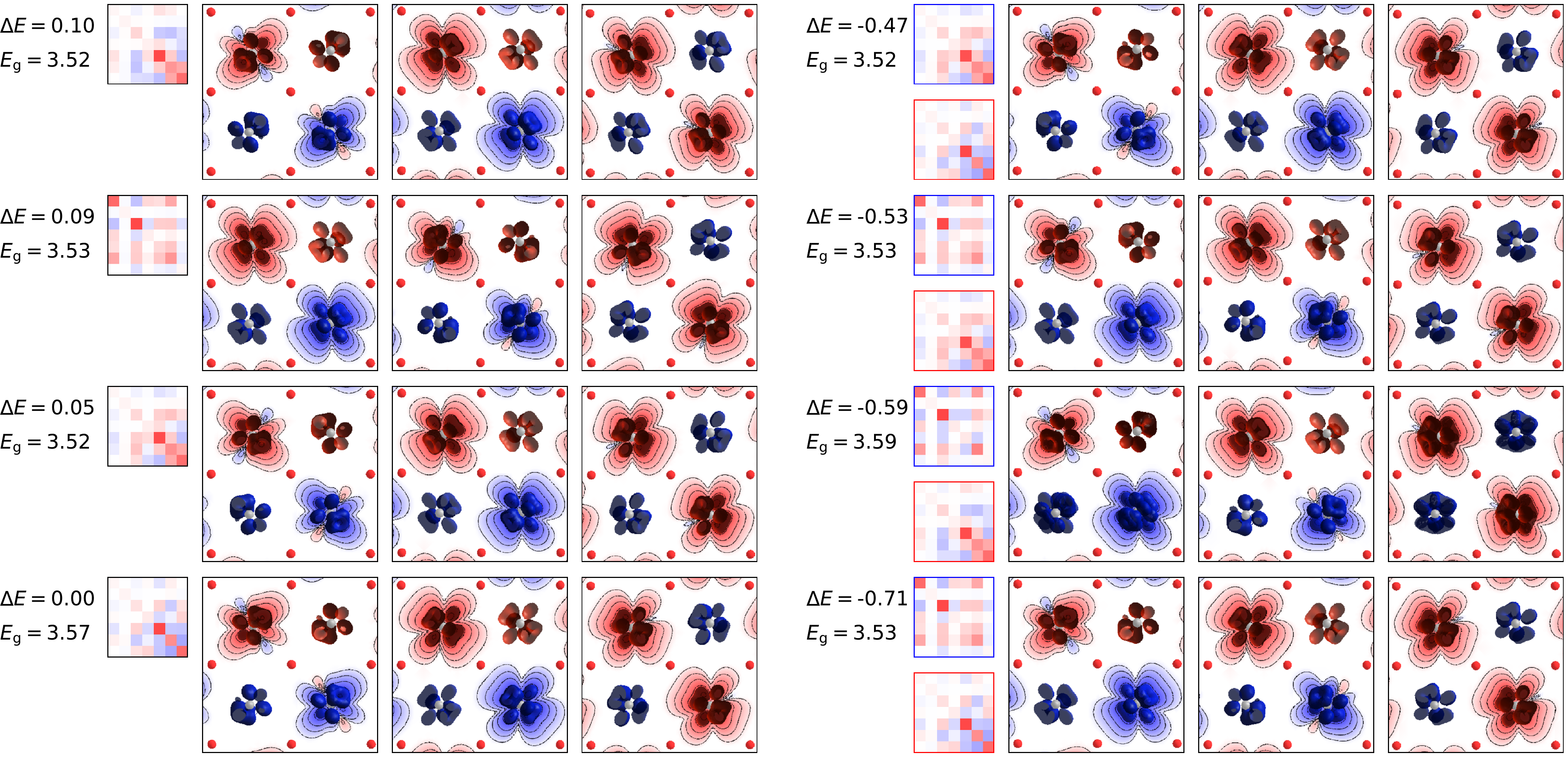}
\caption{The relative energies, $\Delta E$ (meV/atom), band gaps, $E_g$ (eV), OMs and spin densities along the three axes of the four lowest energy states where $O^{\mathrm{UP}} \equiv O^{\mathrm{DOWN}}$ (left) and $O^{\mathrm{UP}} \neq O^{\mathrm{DOWN}}$ (right) respectively, as calculated in a 96 atom cell.  Where they differ, up (down) OMs are indicated by a blue (red) border.  Similarly, blue (red) indicates spin up (down) electron density, while the spin densities have been cropped to show a representative section of the supercell. The depicted isosurfaces correspond to a density of $\pm 0.1$~$e a_0^{-3}$.
}\label{fig:fin}
\end{figure*}

In order to explore in more detail the lowest energy states, in Fig~\ref{fig:fin} we show the OMs, relative energies, band gaps and spin densities for the four lowest energy states both with and without the constraint that $O^{\mathrm{UP}} \equiv O^{\mathrm{DOWN}}$.  Although the OMs, and correspondingly the spin densities show noticeable differences, the band gaps nonetheless differ by less than 0.1~eV.  Regarding the OMs, there are also some preferentially occupied orbitals, as might be expected.  For 
example, $f_{-3}$ and $f_{-1}$ are frequently strongly occupied together.

Finally, we have repeated the overall lowest energy OM setup in a 324 atom supercell.  
The obtained band gap is 3.9~eV, which is significantly higher than the experimental value. 
Although it may be possible that a lower energy state could be found by exploring additional OMs, the difference compared to experiment is large enough that future work should focus on exploring alternative hybrid functionals to PBE0.
For example, Prodan~\emph{et al.}~\cite{Prodan2006} also found that PBE0 overestimates the band gap significantly, while the screened HSE functional~\cite{Heyd2004} gave a value which is much closer to experiment, albeit still an overestimation.  However, the value we find is much higher than Prodan~\emph{et al.}'s value of 3.1~eV, which could be attributed to the difference in basis set or pseudopotentials (effective core potentials). On the other hand it is also possible that they found a higher energy metastable state -- they used a density patching approach to explore different initial charge densities and thereby indirectly control the orbital occupancies, and it is unclear to what extent such an approach is able to explore the complex energy landscape which our calculations have elucidated.  Nonetheless, HSE might prove to be a better choice than PBE0 for future explorations of the energy landscape, while it would be interesting to see to what extent the the energy landscape changes when a different hybrid functional is used.

\subsection{Density of States}

We have also calculated the DoS for both the $2\times 2\times 2$ supercell calculations, which are depicted in the Supplementary Information.  While the $2\times 2\times 2$ supercell calculations are not large enough to be fully converged, it is nonetheless interesting to compare the DoS coming from different metastable states.
In general, the differences between the depicted DoS are small, although there are some subtle variations in the shape of the peak near the valence band edge, which has previously been shown to largely correspond to the U~5$f$ electrons~\cite{Prodan2006}. However, there is no obvious trend relating the nature of the OM and the associated changes in DoS.

\begin{figure}
\centering
\subfigure[DFT] {\includegraphics[scale=0.38]{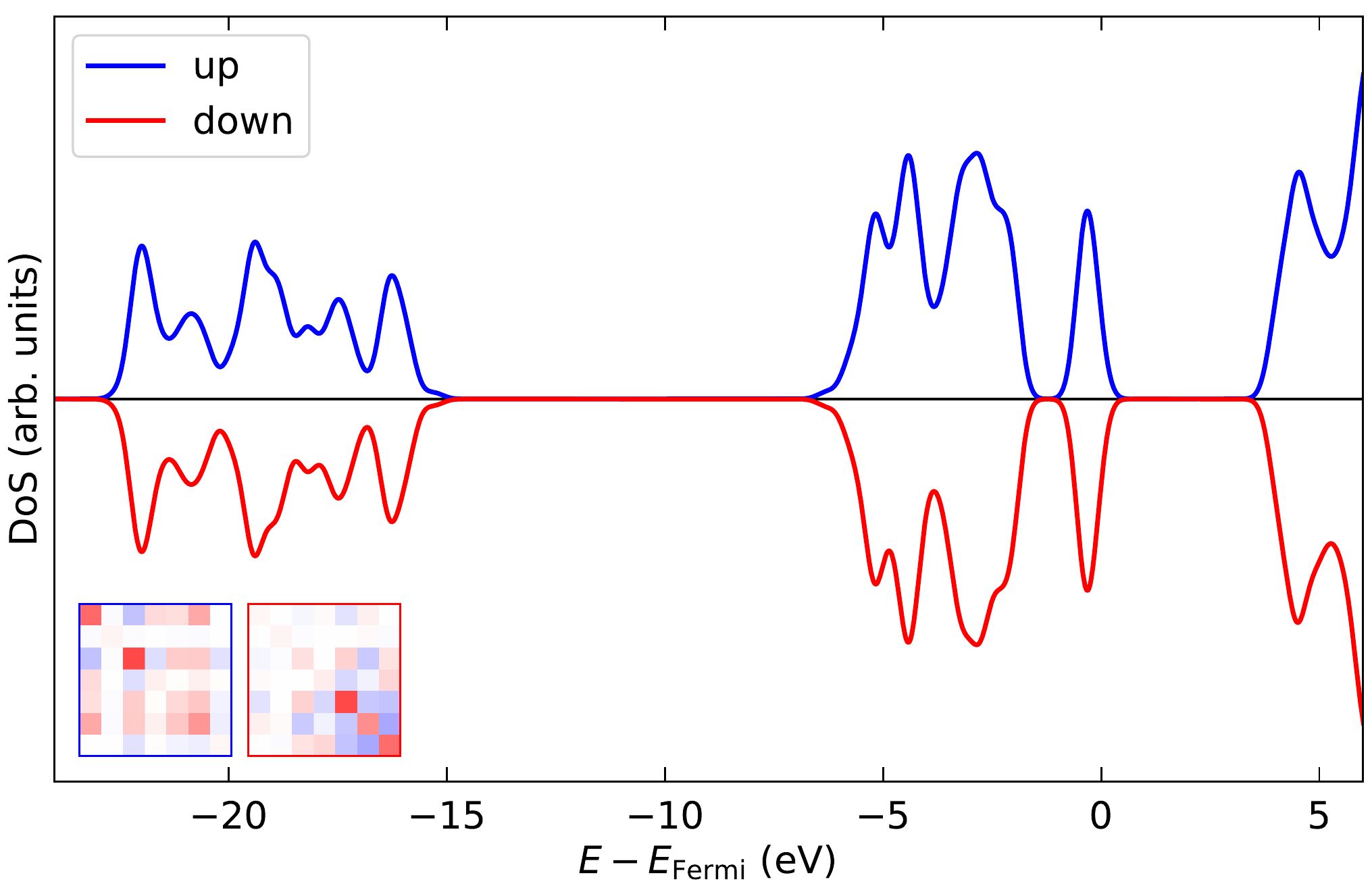}\label{fig:dft_dos}}
\subfigure[DMFT] {\includegraphics[scale=0.38]{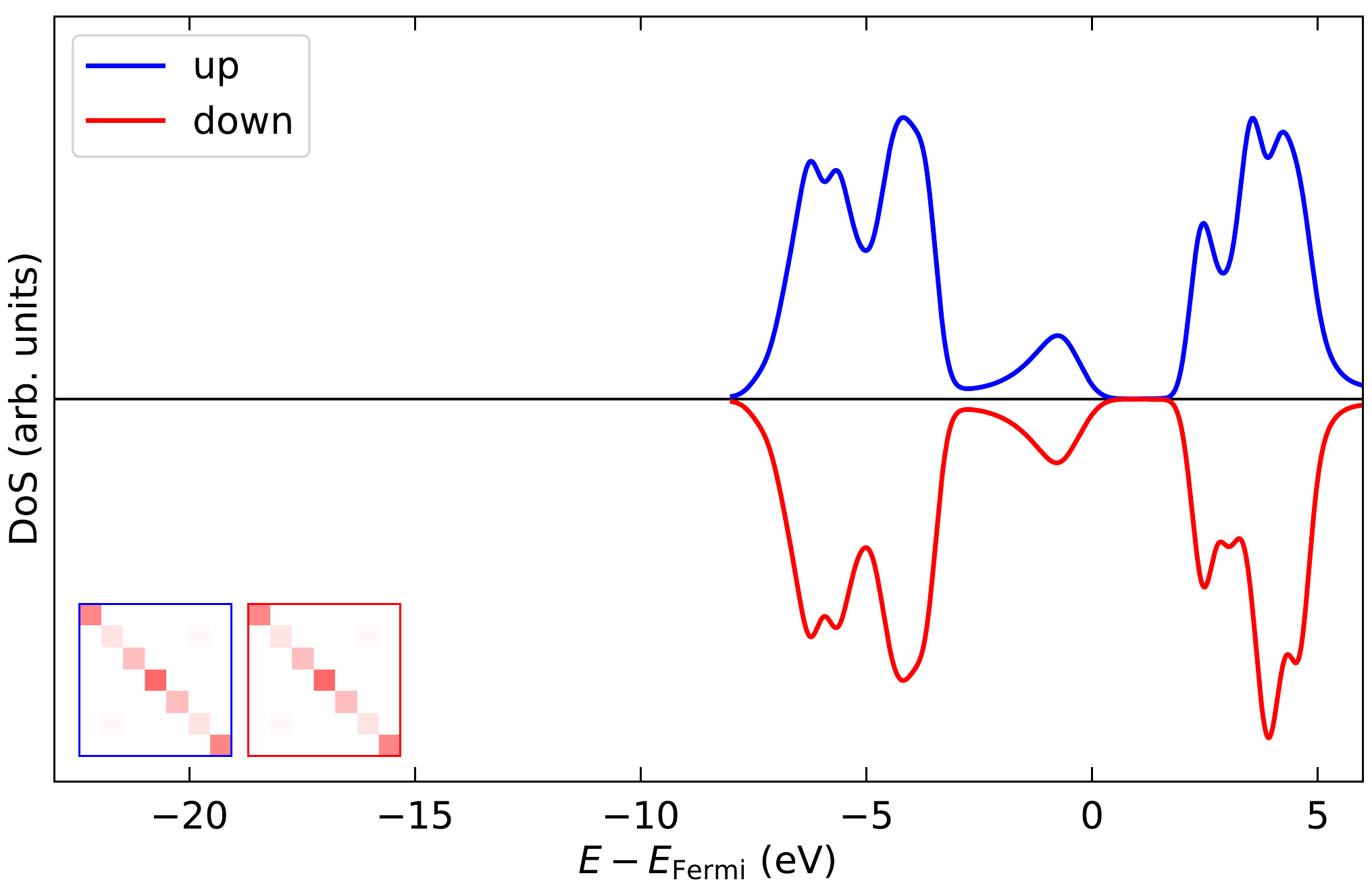}\label{fig:dmft_dos}}
\caption{Densities of states for the final $3\times 3 \times 3$ supercell DFT calculation with PBE0, compared to the DMFT-calculated DoS.  Up and down spin components are depicted, also shown are the OMs corresponding to up (down) spin channels, as indicated by blue (red) borders.
In both cases, Gaussian smearing of 0.2~eV has been applied, while the curves have been shifted so that the Fermi energy, $E_{\mathrm{Fermi}}$, is at zero.
}\label{fig:dos}
\end{figure}

Going to the $3 \times 3 \times 3$ supercell, the DoS for the lowest obtained state is depicted in Fig.~\ref{fig:dft_dos}.
Comparing to available photoemission data from Roy \emph{et al.}~\cite{Roy2008}, the DoS is generally in good agreement.  
Although it is not possible to quantitatively compare the peak heights with experiment without accounting for the differing photoionization cross sections of the different orbitals, the main features of the valence band are nonetheless well described.
The main discrepancy lies in in the separation between the two main peaks in the valence band, which is smaller than in the experimental spectrum.

Finally, we compare to the DMFT results.  Notably, the band gap for DMFT ($\sim$2.0 eV) (measured as the gap between highest occupied molecular orbital (HOMO) and lowest unoccupied molecular orbital (LUMO) band edges), is smaller than for DFT and is much more closely aligned with the experimental value. Comparing the DoS, which is depicted in Fig.~\ref{fig:dmft_dos}, with that of DMFT, we find that the overall shape of the DMFT DoS is quite similar to that of DFT except that the HOMO orbital spectrum in DMFT is much broader due to the dynamical correlation effect. The nature of this HOMO peak below the Fermi energy originates from the Zhang-Rice state due to the charge-transfer type of the insulating state. This Zhang-Rice state due to hybridization between the localized U 5$f^2$ moment the O ligand holes has been discussed in previous DMFT calculations~\cite{Yin2011,Huang2017}.
The positions of DMFT occupied spectral peaks are also consistent with experimental photoemission spectral peaks located at ca.\ -1eV and -5eV~\cite{Roy2008}.

Although the densities of states obtained from DMFT and PBE0 are similar, the OMs coming from DMFT are quite different from those found with PBE0.
In DMFT, the 1-$\mathbf{k}$ AFM structure with 2 U atoms per unit cell was used, with the dynamical fluctuation is treated in a local site while the spatial fluctuation is included in a mean-field way. 
Therefore, the orbital occupation computed in DMFT is highly constrained with higher symmetry suppressing the orbital fluctuation.
In contrast, the spatial disorder effect is treated explicitly in PBE0 within the $3\times3\times3$ supercell although only static correlation is included. 
Furthermore, DMFT used the Wannier functions as a correlated basis for treating the U 5$f$ orbital and the OM basis while PBE0 used the pseudopotential projectors to obtain the OMs.
As a result, mostly diagonal components, particularly $f_{+3}$, $f_{0}$, and $f_{-3}$ are pronounced in DMFT, while $f_{+3}$ and $f_{-3}$ orbitals are almost degenerate.  The OMs in the spin up and down channels in DMFT are similar.
On the other hand, in PBE0, the $f_{-3}$ and $f_{-1}$ orbitals are dominant in the spin up channel while $f_{+3}$ and $f_{+1}$ orbitals are more populated in the spin down channel.
The $f_{-2}$ orbital is energetically not favored, which is different from DMFT.
Indeed, when the DMFT OM was used as an input for PBE0, it was found to be not stable.  It could be interesting in future work to further explore these differences between the two approaches.

\section{Conclusion}\label{sec:conclusion}

In this work we have presented a detailed exploration of the electronic structure of UO$_2$, using the PBE0 hybrid functional as implemented in a systematic wavelet basis code, which allows access to large supercells.  Through the use of an occupancy biasing approach, we explore a large number of metastable states corresponding to different $f$-electron occupancy matrices, including states which have previously been investigated using DFT+U, and new lower energy states which have previously not been explored.  These include both mixed and disordered states, which have significant off-diagonal terms in the OM, and go beyond occupying only 2 $f$-orbitals.  

The results show a rich energy landscape with a large number of different states within a small energy window, but nonetheless with differing band gaps.  Furthermore, when the OMs of up and down U atoms are allowed to differ, further low energy states are also identified, although the decrease in energy is small (less than 1~meV/atom) compared to the case where all OMs are the same.

This work focused on the PBE0 functional.  While the DoS corresponding to the identified state with the lowest energy agrees well with experimental photoemission data, the band gap is significantly overestimated compared to experiment.  It would therefore be interesting to perform similarly detailed investigations with other hybrid functionals in future, such as HSE, which has been shown to give a more accurate band gap, to see if a similarly rich energy landscape may be found.  

Aside from considerations relating to the functional, it would also be interesting in future to explore states where the OM is allowed to vary across \emph{all} U atoms in the system, rather than just between up and down U atoms.  This might allow for example the ability to investigate spin-wave like states.

We also performed DFT+DMFT calculation of UO$_2$ and compared to the PBE0 results. Although the DMFT DoS shows similar features to the PBE0 DoS, consistent with the experimental data, the OM in DMFT is noticeably different from that in PBE0. Our results show that dynamical correlations can play a distinct role in predicting the distribution of OMs in metastable correlated materials compared to the methods treating only static correlations. Studying the effects of the multi-site treatment in the supercell structure or the choice of correlated basis (Wannier \emph{vs.}\ projector) in DMFT would therefore also be an important subject for future studies.

\section*{Acknowledgements}

LER acknowledges an EPSRC Early Career Research Fellowship (EP/P033253/1) and the Thomas Young Centre under grant number TYC-101.
An award of computer time was provided by the Innovative and Novel Computational Impact on Theory and Experiment (INCITE) program. This research used resources of the Argonne Leadership Computing Facility, which is a DOE Office of Science User Facility supported under Contract DE-AC02-06CH11357.
We are also grateful to the UK Materials and Molecular Modelling Hub for computational resources, which is partially funded by EPSRC (EP/P020194/1), while calculations were also performed on the Imperial College High Performance Computing Service and the ARCHER UK National Supercomputing Service.
HP acknowledges funding from the US Department of Energy, Office of Science, Basic Energy Sciences Division of Materials Sciences and Engineering.
HP gratefully acknowledges the computing resources provided on Bebop, high-performance computing clusters operated by the Laboratory Computing Resource Center at Argonne National Laboratory. Los Alamos National Laboratory is managed by Triad National Security, LLC, for the National Nuclear Security Administration of the U.S. Department of Energy under Contract No.\ 89233218CNA000001.

\section*{References}

\bibliographystyle{iopart-num}
\bibliography{uo2_refs}

\end{document}


\title{\LARGE Supplementary Information\\
\vspace{5pt}
\large Exploring Metastable States in UO$_2$ using Hybrid Functionals and Dynamical Mean Field Theory}

\author{Laura E.\ Ratcliff}     \affiliation{\ICL}
\author{Luigi Genovese}         \affiliation{\CEA}
\author{Hyowon Park}\affiliation{\ANL}\affiliation{\UIC}
\author{Peter B. Littlewood}     \affiliation{\ANL}\affiliation{\UofC} 
\author{Alejandro Lopez-Bezanilla}    \affiliation{\LANL}

\date{\today}

\beginsupplement

\maketitle

\section{Basis Set Convergence}

\begin{figure}[!h]
\centering
\includegraphics[height=0.49\textwidth,angle=-90]{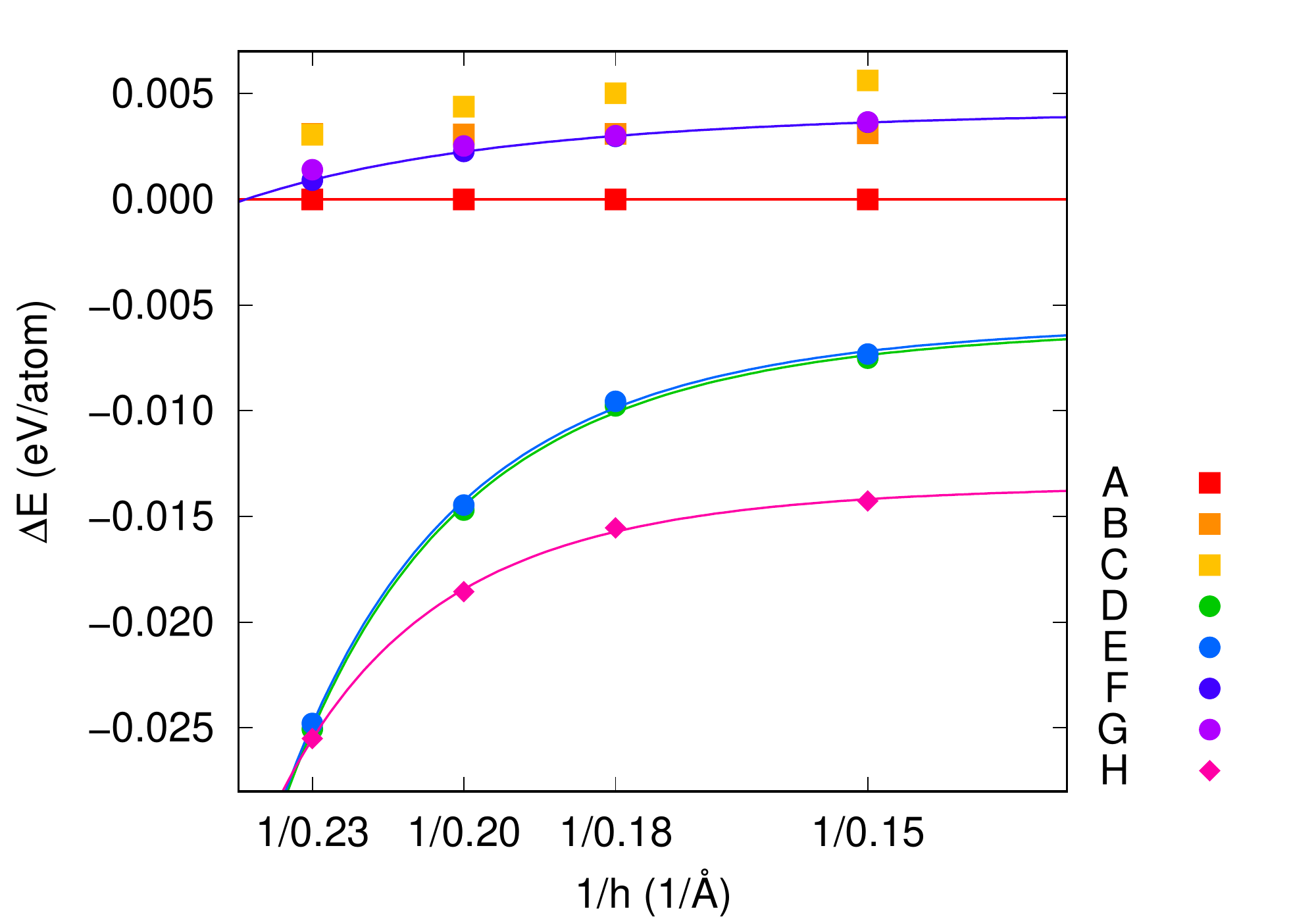}
\caption{Convergence of relative energies with respect to the energy of state A at a given grid spacing, \emph{vs.}\ inverse grid spacing, $h$, in \AA\  (i.e.\ increasing basis set size), for selected input occupancies in a 96 atom cell. The different symbols refer to whether the state is pure, mixed, or disordered. Where appropriate, an exponential function has been fitted (lines), to facilitate extrapolation to a complete basis.}\label{fig:hgrid_conv}
\end{figure}

In order to explore the convergence with respect to the wavelet grid spacing, a representative selection of states (labelled A-H)  were taken, neglecting high energy states and those which are degenerate due to symmetry.
Out of the disordered states, state H was selected for further investigation, since it is the only one of comparable energy to the lowest energy mixed states.  Since the number of SCF iterations required to converge this state was extremely high, the wavefunctions from the initial calculation in the 96 atom cell were used as an input for subsequent calculations, where an interpolation scheme~\cite{Ratcliff2015} was used to adjust the Kohn Sham wavefunctions for either a change in cell size or in the grid spacing of the wavelet basis set.
The final OMs of states D, E and F were directly imposed during the occupancy biasing stage, since the number of iterations required to converge was also relatively high.  
Fig.~\ref{fig:hgrid_conv} and Table~\ref{tab:energies_hgrid} show the convergence of the relative energies and band gaps for states A through H for decreasing grid spacing, $h$.

\begin{table}
\centering
\begin{tabular}{c rr c cc cc c cc c cc c cc c cc}
\toprule
 & \multicolumn{2}{c}{Input} & Final &&& \multicolumn{2}{c}{$h=0.23$} && \multicolumn{2}{c}{$h=0.20$}  && \multicolumn{2}{c}{$h=0.18$} && \multicolumn{2}{c}{$h=0.15$} && \multicolumn{2}{c}{$h\rightarrow0$}\\
& $f_i$& $f_j$  & OM &&& $\Delta E$ & E$_{\mathrm{g}}$ && $\Delta E$ &  E$_{\mathrm{g}}$ &&  $ \Delta E$ & E$_{\mathrm{g}}$ && $\Delta E$ &  E$_{\mathrm{g}}$ && $\Delta E$ &  E$_{\mathrm{gp}}$\\
\cmidrule{1-4}   \cmidrule{7-8} \cmidrule{10-11} \cmidrule{13-14} \cmidrule{16-17} \cmidrule{19-20}\\[-2.5ex]

A & +1 & +3  
&\begin{minipage}{.05\textwidth}
      \includegraphics[width=\linewidth]{om2.png}
    \end{minipage}
&&&  0.000 & 2.6 && 0.000 & 2.7 &&  0.000 & 2.7 &&
 0.000 &  2.7  && - & - \\
\cmidrule{1-4}   \cmidrule{7-8} \cmidrule{10-11} \cmidrule{13-14} \cmidrule{16-17} \cmidrule{19-20}\\[-2.5ex]

B & 0 & +2 
& \begin{minipage}{.05\textwidth}
      \includegraphics[width=\linewidth]{om3.png}
    \end{minipage}
&&& 0.003 &  2.9 &&  0.003 &  2.9  && 0.003 & 2.9 &&
0.003 &  2.9 &&  - &  - \\
\cmidrule{1-4}   \cmidrule{7-8} \cmidrule{10-11} \cmidrule{13-14} \cmidrule{16-17} \cmidrule{19-20}\\[-2.5ex]

C & -1 &  +3
& \begin{minipage}{.05\textwidth}
      \includegraphics[width=\linewidth]{om5.png} 
    \end{minipage}
 &&& 0.003 & 2.9 && 0.004 & 3.0  &&  0.005 & 2.9&& 
 0.006 & 2.9 &&  - &  - \\

\midrule \\[-3.5ex]
\midrule \\[-2.5ex]
    
\multirow{2}{*}{\vspace{-5pt} D} & -3 &  +3 
& \multirow{2}{*}{\begin{minipage}{.05\textwidth}
      \includegraphics[width=\linewidth]{om8.png}
    \end{minipage}}\vspace{5pt}
&&& \multirow{2}{*}{\vspace{-5pt} -0.025} & \multirow{2}{*}{\vspace{-5pt}  3.4}  && \multirow{2}{*}{\vspace{-5pt}  -0.015} & \multirow{2}{*}{\vspace{-5pt}  3.4} && \multirow{2}{*}{\vspace{-5pt} -0.010} & \multirow{2}{*}{\vspace{-5pt} 3.4} &&  
 \multirow{2}{*}{\vspace{-5pt} -0.008} & \multirow{2}{*}{\vspace{-5pt} 3.3} &&\multirow{2}{*}{\vspace{-5pt} -0.006} & \multirow{2}{*}{\vspace{-5pt} -} \\
& -1 &  +1 & &&& & && & &&   & &&
 &  &&  &    \\
\cmidrule{1-4}   \cmidrule{7-8} \cmidrule{10-11} \cmidrule{13-14} \cmidrule{16-17} \cmidrule{19-20}\\[-2.5ex]

\multirow{2}{*}{\vspace{-5pt} E} & -3 &  +2 
& \multirow{2}{*}{\begin{minipage}{.05\textwidth}
      \includegraphics[width=\linewidth]{om10.png} 
    \end{minipage}}\vspace{5pt}
&&& \multirow{2}{*}{\vspace{-5pt} -0.025} & \multirow{2}{*}{\vspace{-5pt} 3.1} && \multirow{2}{*}{\vspace{-5pt} -0.014} & \multirow{2}{*}{\vspace{-5pt} 3.1} && \multirow{2}{*}{\vspace{-5pt} -0.010} & \multirow{2}{*}{\vspace{-5pt}3.1} &&
 \multirow{2}{*}{\vspace{-5pt} -0.007} & \multirow{2}{*}{\vspace{-5pt} 3.1} && \multirow{2}{*}{\vspace{-5pt} -0.006} & \multirow{2}{*}{\vspace{-5pt} -} \\
& -1 &  +2 & &&& \\
\cmidrule{1-4}   \cmidrule{7-8} \cmidrule{10-11} \cmidrule{13-14} \cmidrule{16-17} \cmidrule{19-20}\\[-2.5ex]

F & -3 &  0 
& \begin{minipage}{.05\textwidth}
      \includegraphics[width=\linewidth]{om14.png}
    \end{minipage}
&&& 0.001 &3.2 &&  0.002 & 3.2&&0.003 & 3.2&&
0.004 & 3.2 && 0.004 &  - \\
\cmidrule{1-4}   \cmidrule{7-8} \cmidrule{10-11} \cmidrule{13-14} \cmidrule{16-17} \cmidrule{19-20}\\[-2.5ex]

G & -1 &  0 
& \begin{minipage}{.05\textwidth}
      \includegraphics[width=\linewidth]{om16.png}
    \end{minipage}
&&&  0.001 & 2.8 && 0.003 &  2.8 && 0.003 & 2.8 &&
0.004 & 2.8 && - & - \\

\midrule \\[-3.5ex]
\midrule \\[-2.5ex]
    
H & -3 &  -2 
& \begin{minipage}{.05\textwidth}
      \includegraphics[width=\linewidth]{om18.png} 
    \end{minipage}
&&& { -0.026} & { 3.2} && {-0.019} & { 3.2} && -0.016 & 3.2 &&
-0.014 & 3.2 && -0.013 &  -\\
  \bottomrule
\end{tabular}
\caption{Convergence with respect to grid spacing of relative energies, $\Delta E$, (with respect to state A at that grid spacing) (eV/atom), band gaps, $E_{\mathrm{g}}$ (eV) and average OMs for selected input occupancies in a 96 atom cell. 
Where an exponential curve could be fitted the extrapolated value for relative energy in a complete basis set is also given.}  
\label{tab:energies_hgrid}
\end{table}

\clearpage

\section{Lattice Constant}

\begin{figure}[!h]
\centering
\includegraphics[height=0.49\textwidth,angle=-90]{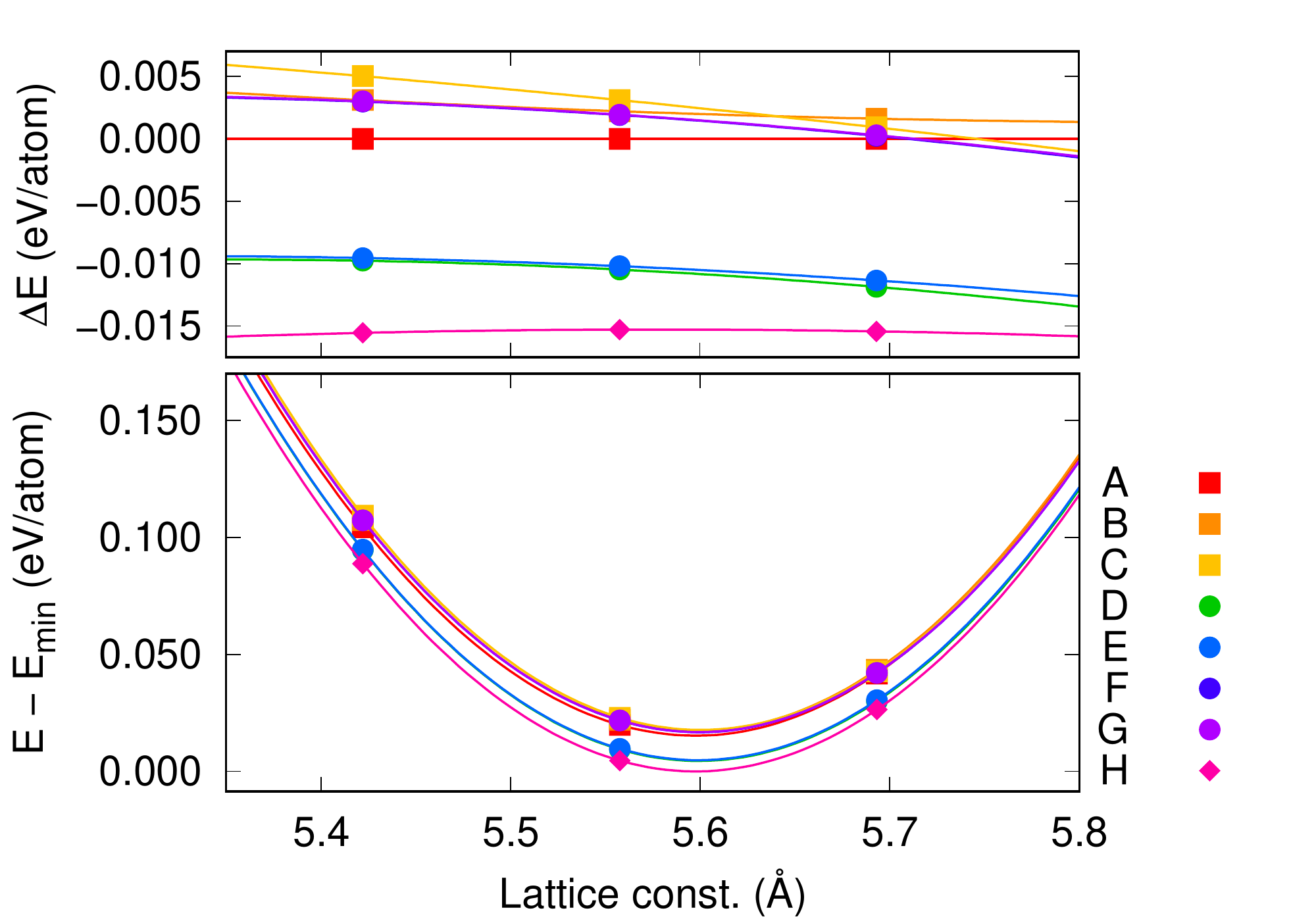}
\caption{Variation in total energy [bottom] and relative energies [top] with respect to the cubic lattice constant for $2 \times 2 \times 2$ supercells.  The curves on the bottom panel have been vertically shifted for easier comparison, while the different symbols refer to whether the state is pure, mixed, or disordered. }\label{fig:latt}
\end{figure}

Fig.~\ref{fig:latt} shows both the energies of states A-H with respect to the lattice constant and the corresponding variation in relative energies for a 96 atom supercell using a grid spacing of $0.18$~\AA.
For all states, a quadratic polynomial was fitted from three calculations; in order to verify that this was sufficient to find the equilibrium lattice constant, a further two points were generated for state D, which had negligible impact on the fitted curve and thereby the obtained lattice constant.
For state D the simulations were also repeated for a larger 324 atom supercell, as shown in the main text.
The wavelet grid spacing was found to have a significant impact on the determination of the optimal lattice constant, and so it proved to be important to properly converge the basis before optimizing the cell size.

\clearpage

\section{Exploring the Occupancy Space}

\begin{figure*}[!h]
\centering
\subfigure[$\Delta E$]{\includegraphics[width=0.35\textwidth]{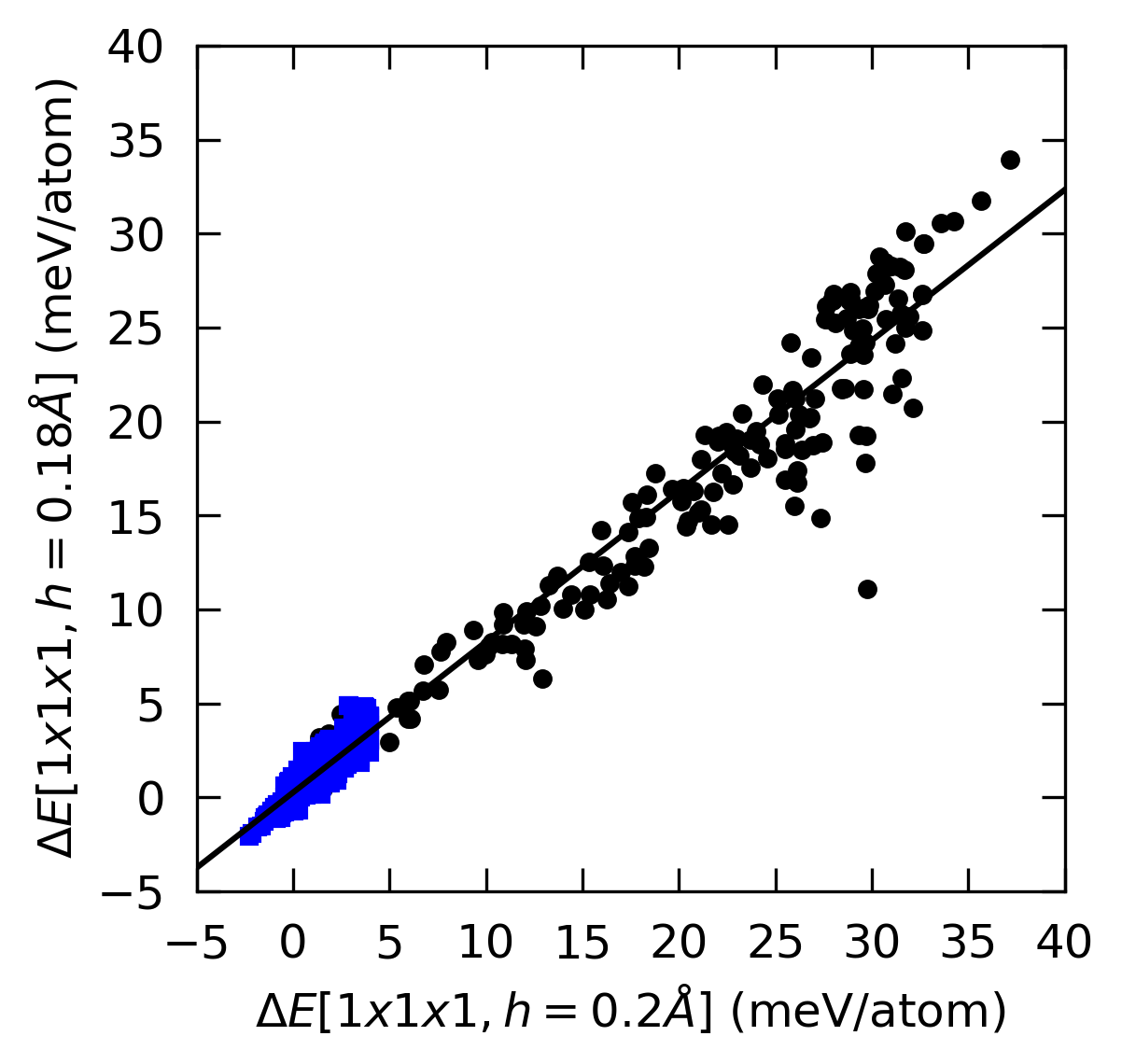}}
\subfigure[$E_{\mathrm{g}}$]{\includegraphics[width=0.35\textwidth]{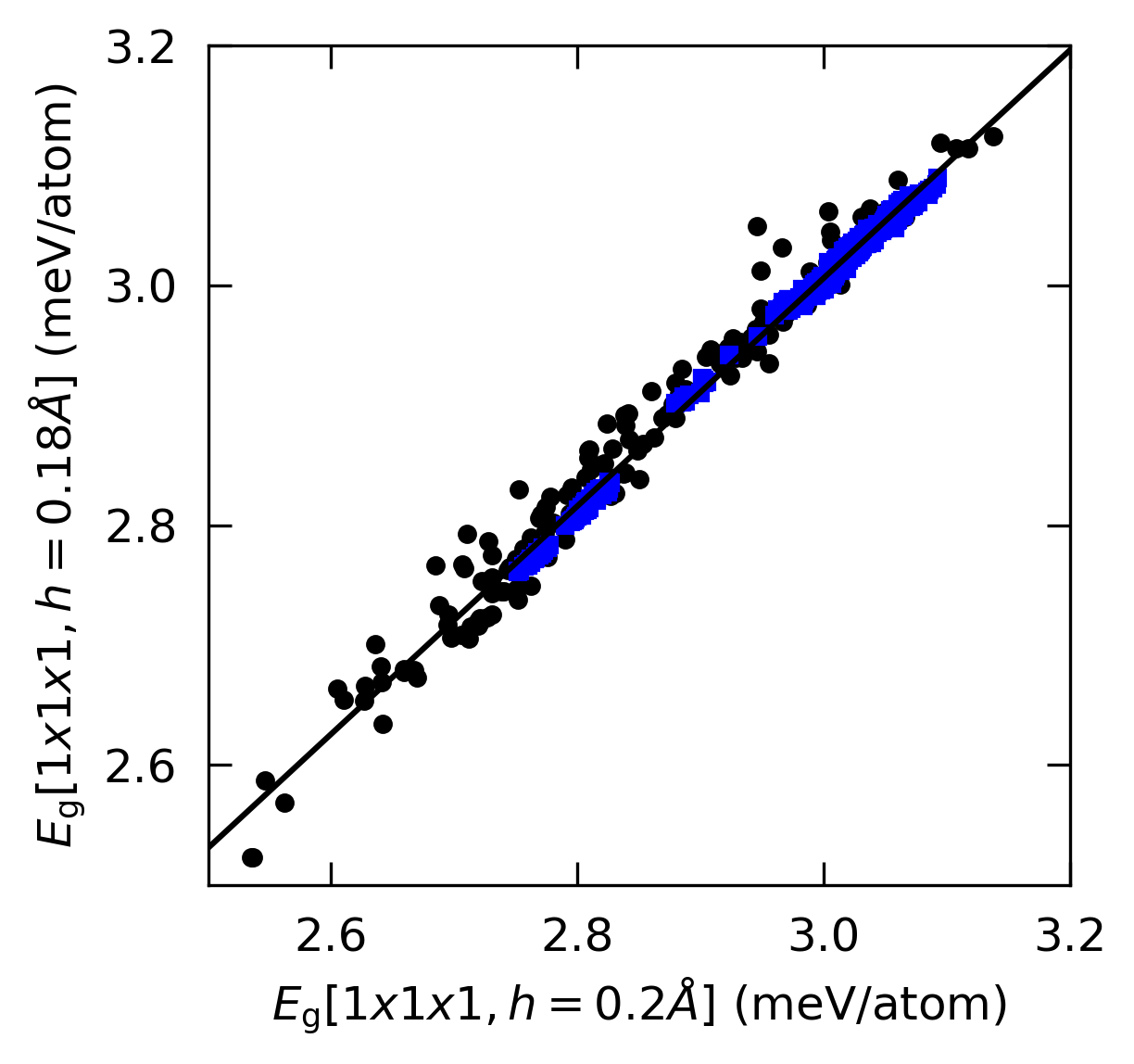}}\\
\subfigure[$\Delta E$]{\includegraphics[width=0.35\textwidth]{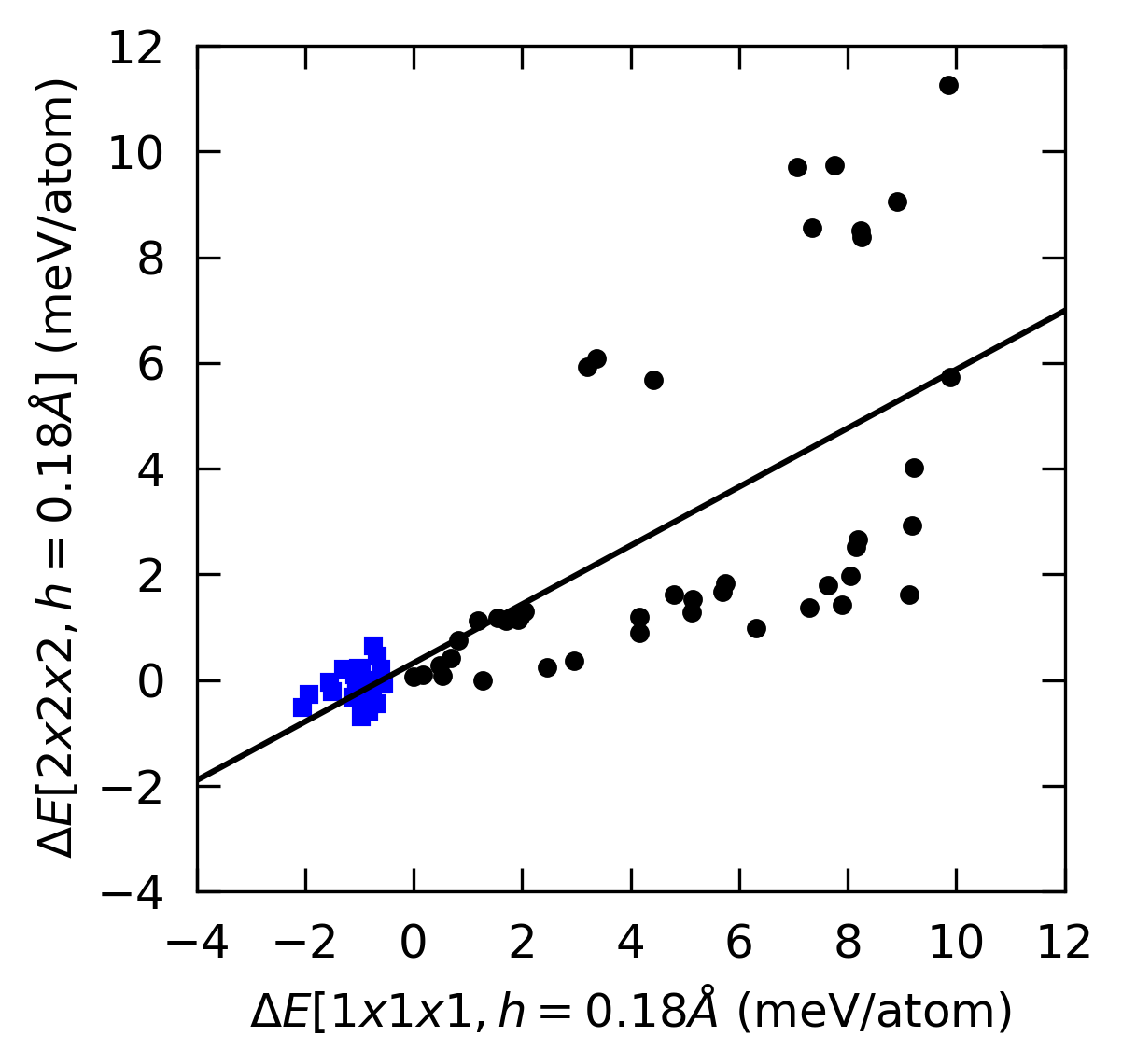}}
\subfigure[$E_{\mathrm{g}}$]{\includegraphics[width=0.35\textwidth]{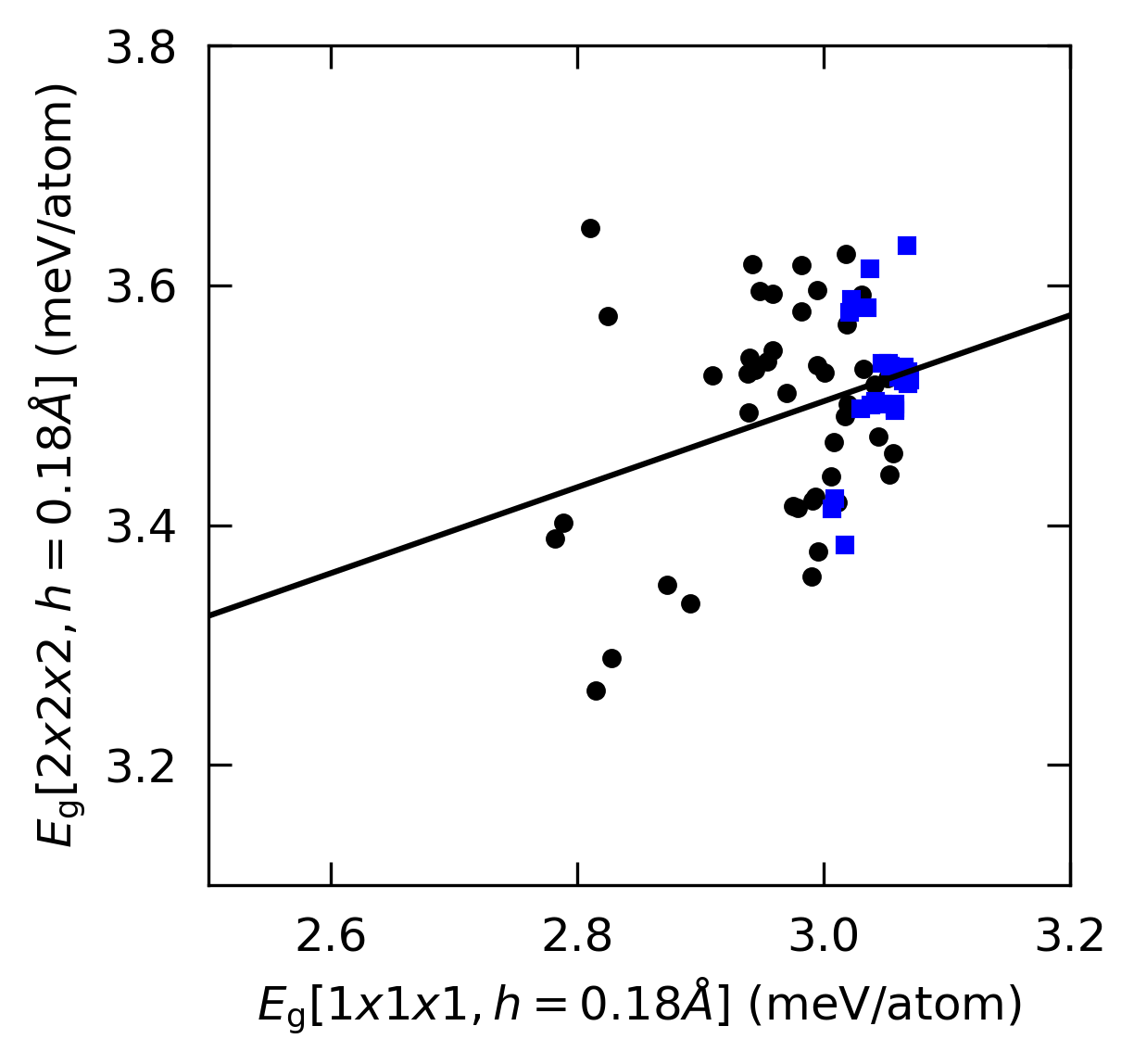}}
\caption{Correlation between results from different calculation setups for relative energies ($\Delta E$) [left] and band gaps ($E_{\mathrm{g}}$) [right], when varying the grid spacing in a small cell [top] and when changing the supercell size at a fixed grid spacing [bottom].  The black circles are for the case where the imposed OM is the same for all atoms in the system, while the blue squares denote calculations where up and down U atoms have different imposed OMs.  The straight lines represent a linear fit.}\label{fig:uo2_s111_v_s222}
\end{figure*}

\begin{figure*}[!h]
\centering
\subfigure[$1\times 1\times 1$]{\includegraphics[width=0.35\textwidth]{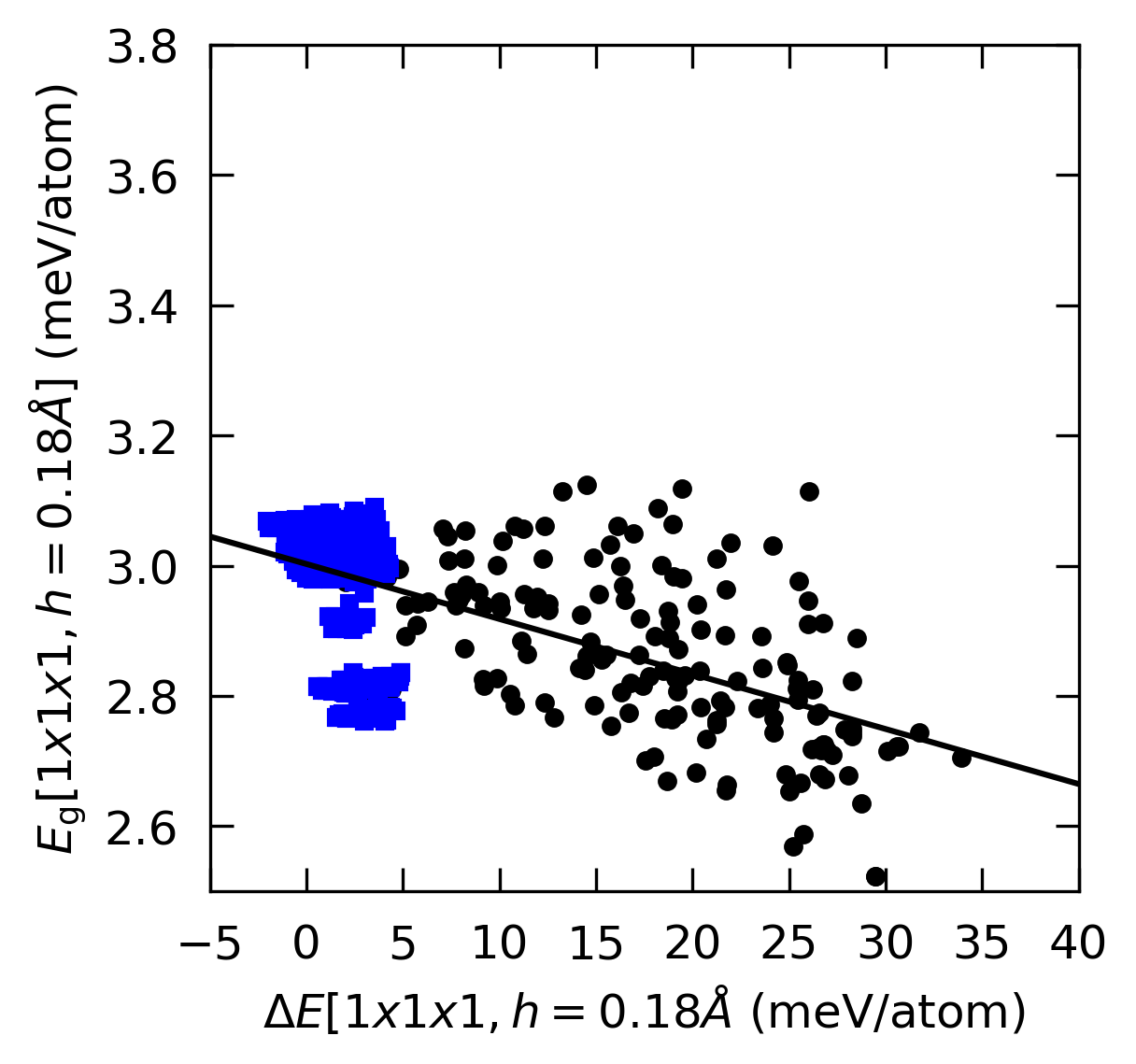}}
\subfigure[$2\times 2\times 2$]{\includegraphics[width=0.35\textwidth]{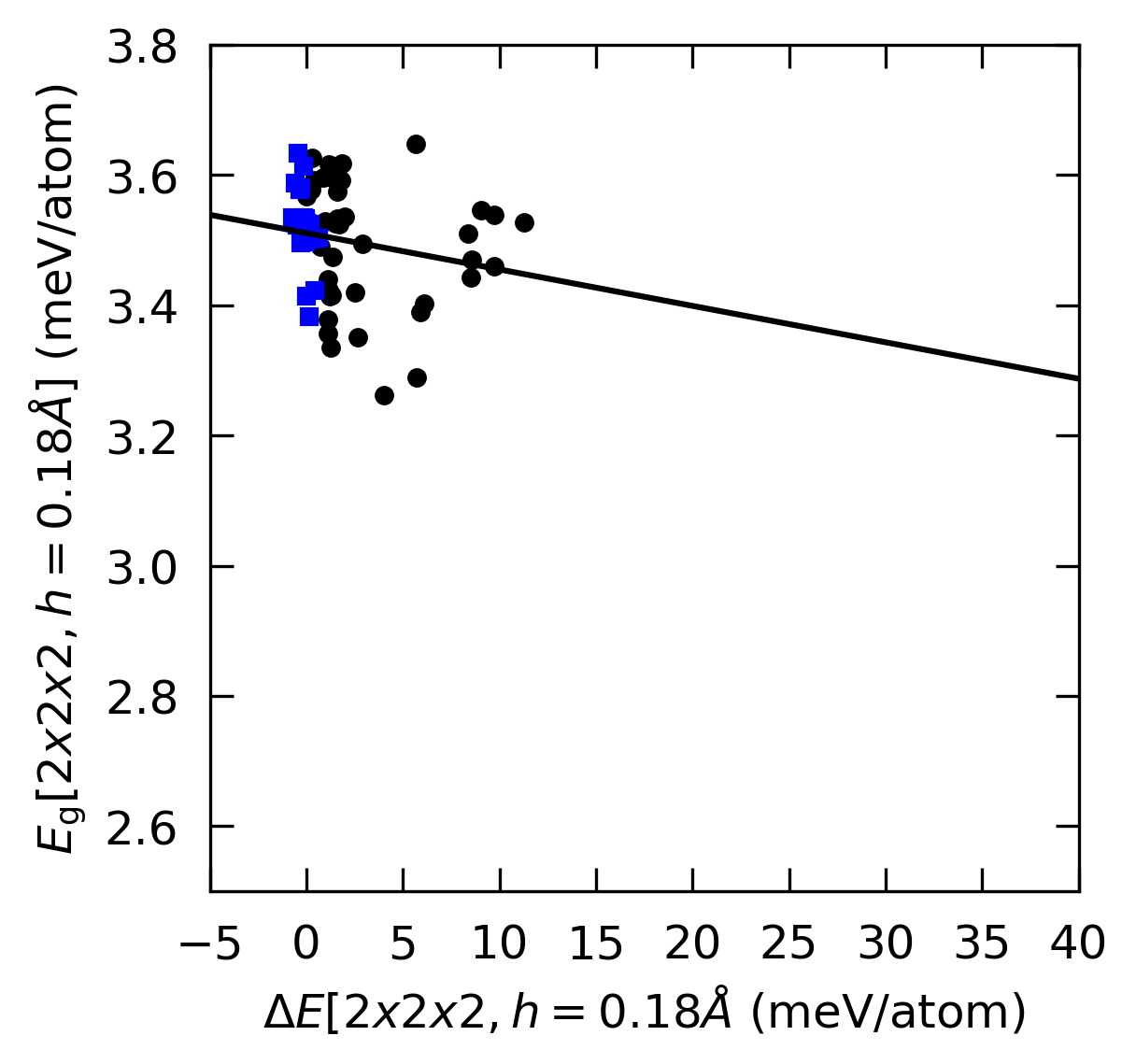}}
\caption{Correlation between relative energies ($\Delta E$) and band gaps ($E_{\mathrm{g}}$) for different supercell sizes.  The black circles are for the case where the imposed OM is the same for all atoms in the system, while the blue squares denote calculations where up and down U atoms have different imposed OMs. The straight lines represent a linear fit.}\label{fig:uo2_e_v_g}
\end{figure*}

\clearpage

\section{Densities of States}

\begin{figure}[!h]
\centering
\includegraphics[scale=0.39]{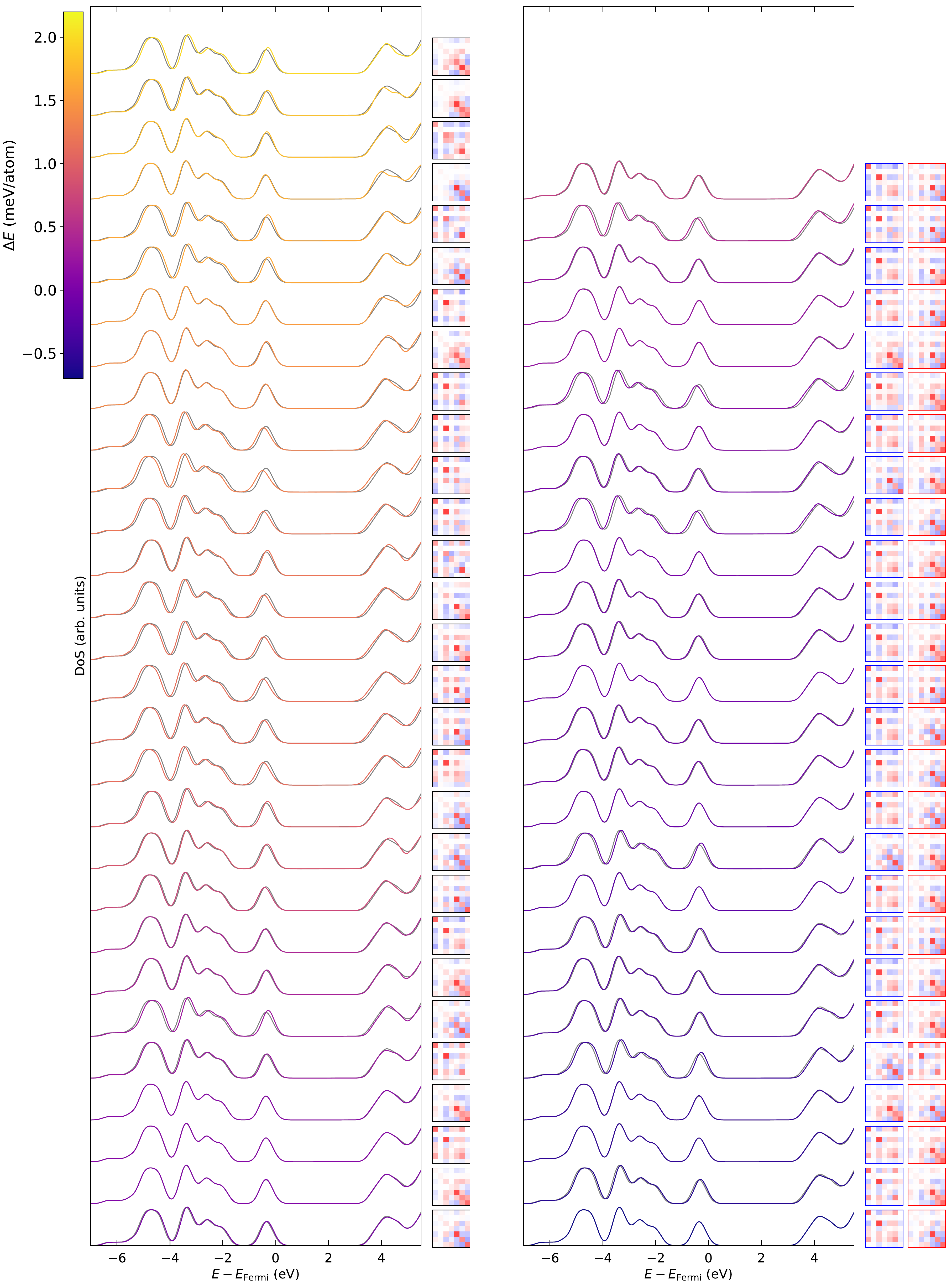}
\caption{Spin up densities of states for all 96 atom calculations where $O^{\mathrm{UP}} \equiv O^{\mathrm{DOWN}}$ [left] and $O^{\mathrm{UP}} \neq O^{\mathrm{DOWN}}$ [right], where the colour denotes the relative energy.  Where they are distinct, up (down) OMs are depicted with a blue (red) border. Gaussian smearing of 0.2~eV has been applied, while the curves have been shifted so that the Fermi level, $E_{\mathrm{Fermi}}$, is at zero.  The DoS for the lowest energy state is superimposed in grey for comparison.}\label{fig:dos}
\end{figure}

\bibliographystyle{apsrev4-1}
\bibliography{uo2_refs}